\PassOptionsToPackage{svgnames}{xcolor}
\documentclass[12pt]{article}

\usepackage[a4paper]{geometry} % Geometry for setting page margins
\usepackage{amsmath, amssymb, amsfonts, amsthm, setspace, tikz, dsfont, wrapfig, lipsum, natbib, array, colortbl, mathrsfs, pifont, wasysym, nicefrac, multicol, booktabs, makecell, multirow, caption, subcaption, accents, graphicx, comment}
\usepackage{algorithm}
\usepackage{algpseudocode}
\usepackage{svg, pgfplots, etoolbox, thmtools, thm-restate, enumerate, centernot, fancyhdr, ifthen, hyperref, cleveref, cancel, bbm, refcount}
\usepackage{chngcntr}
\usepackage{ragged2e}
\usetikzlibrary{patterns, arrows.meta, positioning, shapes.geometric, plotmarks, pgfplots.colormaps}
\pgfplotsset{compat=1.18, plot coordinates/math parser=false}
\usepgfplotslibrary{patchplots, colormaps, groupplots}

\hypersetup{
    pdffitwindow=true, % Fit PDF to window
    pdfstartview={XYZ null null 1.00}, % Start with 100% zoom
    pdfnewwindow=true, % Open links in new window
    colorlinks=true, % Color links instead of using boxes
    linkcolor=DarkBlue, % Color for internal links
    citecolor=DarkBlue, % Color for citations
    urlcolor=DarkBlue, % Color for external URLs
    pdfauthor={Filip Obradović and Gabriel Ziegler}, % Author metadata
    pdfkeywords={}, % Keywords metadata
    pdftitle={Knowing Less After a Diagnostic Test: Dilation in Clinical Decision-Making}, % Title metadata
    pdfsubject={}, % Subject metadata
    pdfpagemode=UseNone % Don't use any page mode when opening
}

\newtheorem{definition}{Definition}
\newtheorem{assumption}{Assumption}

\newtheorem{proposition}{Proposition}
\newtheorem{theorem}{Theorem}
\newtheorem{corollary}{Corollary}
\newtheorem{remark}{Remark}

\newenvironment{assumptionS}[1]
    {\renewcommand{\theassumption}{\getrefnumber{#1}S\relax}\begin{assumption}\addtocounter{assumption}{-1}}
    {\end{assumption}}

\theoremstyle{definition}

\fancypagestyle{appendix}{
    \fancyhead{}
    \fancyhead[R]{\ifthenelse{\isodd{\value{page}}}{\thepage}{}}
    \fancyhead[L]{\ifthenelse{\isodd{\value{page}}}{\textsc{}}{\thepage}}
    
    \fancyfoot{}
}

\DeclareMathOperator{\inter}{int}

\newcommand{\independent}{\perp\mkern-9.5mu\perp} % Independence symbol
\newcommand{\notindependent}{\centernot{\independent}} % Not independent symbol

\usepackage{titlesec}
\titleformat*{\section}{\large\scshape\centering}
\titleformat*{\subsection}{\scshape\centering}
\titleformat*{\subsubsection}{\itshape}
\titleformat*{\paragraph}{\large\bfseries\centering}
\titleformat*{\subparagraph}{\large\bfseries\centering}

\makeatletter
\DeclareRobustCommand\citepos{
    \begingroup
    \let\NAT@nmfmt\NAT@posfmt
    \NAT@swafalse\let\NAT@ctype\z@\NAT@partrue
    \@ifstar{\NAT@fulltrue\NAT@citetp}{\NAT@fullfalse\NAT@citetp}
}
\let\NAT@orig@nmfmt\NAT@nmfmt
\def\NAT@posfmt#1{\NAT@orig@nmfmt{#1's}}
\makeatother

\title{
    \vspace{-1em}
    Knowing Less After a Diagnostic Test: Dilation in Clinical Decision-Making\thanks{
     This paper supersedes \cite{ziegler-2021}. We are thankful for comments and questions from Charles Manski, Ivan Canay, Federico Bugni, Eric Mbakop and audience members at the Pennsylvania Theory Conference 2023, the Durham York Theory Workshop 2023, the Theory Day 2023 at the University of Edinburgh, the NOeG Winter Workshop 2023, the 24th Annual SAET Conference 2025, and seminar participants at the University of Graz, NYU Shanghai, Royal Holloway University of London, UC Berkeley, UC Davis, University of Toronto, University of Surrey, University of Udine, Yale University, the Berlin Micro Theory Seminar Series, and the Berlin Behavioral Economics Group. Parts of the research were conducted while Ziegler was visiting the University of Pittsburgh and NYU Shanghai, and he expresses his gratitude for their hospitality. Obradovi{\'c} gratefully acknowledges financial support from the Robert Eisner Memorial Fellowship. 
    }
}

\author{
    Filip Obradovi\'{c}\thanks{
        UCLA; \href{mailto:obradovicfilip@ucla.edu}{\tt obradovicfilip@ucla.edu}.
    }
    \and
    Gabriel Ziegler\thanks{
        Freie Universit\"at Berlin and Berlin School of Economics; \href{mailto:gabriel.ziegler@fu-berlin.de}{\tt gabriel.ziegler@fu-berlin.de}.
    }   
}

\date{\today\\
}
\begin{document}

\maketitle
\thispagestyle{empty} % Empty first-page header and footer
\vspace*{-2em} % Adjust spacing after title
\begin{abstract}
\small
Dilation \citep{seidenfeld-wasserman-1993} occurs when conditioning on new information \textit{only} increases uncertainty regardless of what the information reveals. Often viewed as an undesirable feature of models of ambiguity, dilation has sparked sustained debate about belief updating. This paper shows that following established clinical guidance can induce dilation, even for diagnostic tests that meet conventional performance benchmarks. We characterize when a test is dilation-inducing and show that a ubiquitous feature of performance studies -- evaluating tests against an imperfect reference -- creates scope for dilation. To facilitate detection before clinical use and inform regulatory standards, we develop a finite-sample valid hypothesis test of the null that a diagnostic test is dilation-inducing in this context. Applying the method to an influential study that preceded and contributed to the clinical adoption of chest CT screening for COVID-19, we find that the data are consistent with dilation.

\end{abstract}


\cleardoublepage % Ensure the next page starts on a clean page
\setcounter{page}{1} % Reset page numbering

\setstretch{1.3} % Set line spacing again in case of adjustments

\section{Introduction}

Economic theory typically treats additional information as valuable -- more data should (weakly) help an agent determine the state of the world. Under ambiguity, however, it is possible that conditioning on a signal \textit{only} increases uncertainty\footnote{Here, we use ``uncertainty'' broadly, not only in the narrow probabilistic sense.
} regardless of what the signal reveals. This phenomenon, known as \textit{dilation} \citep{seidenfeld-wasserman-1993}, occurs when conditioning on \textit{any} possible signal realization replaces the initial probability assigned to an event with a nondegenerate range of probabilities containing the original value.\footnote{
The authors give the following example on p.~1140: ``\textit{To emphasize the counterintuitive nature of \textit{dilation}, imagine that a physician tells you that you have probability \( \frac{1}{2} \) that you have a fatal disease. He then informs you that he will carry out a blood test tomorrow. Regardless of the outcome of the test, if he conditions on the new evidence, he will then have lower probability 0 and upper probability 1 that you have the disease. Should you allow the test to be performed? Is it rational to pay a fee not to perform the test?''}} 

In standard models of ambiguity, merely updating on a dilation-inducing signal can make an ambiguity-averse agent worse off, even when the information is costless. Dilation is thus often viewed as an undesirable model feature. Its normative implications have generated significant debate \citep{pedersen2014demystifying, gong2021judicious}, and recent work in economic theory develops updating rules explicitly designed to preclude it \citep{gul2021evaluating, tang2022theory}. \citet{shishkin2023ambiguous} document experimental behavior inconsistent with standard models' predictions under dilation.

We show that, despite these concerns, following the standard procedure for using diagnostic tests prescribed by clinical guidelines and guidance can induce dilation, even for tests that meet conventional performance benchmarks. Clinicians are directed to base decisions on the probability of illness conditional on the observed test result (for examples, see \citealp{mcneil1975primer}, \citealp{watson-et-al-2020}, \citealp{ahrq2022probability}, \citealp{vrints2024esc}). To do so, they first form a \textit{pre-test} probability that the patient has a suspected illness based on expert knowledge, prevalence rates, or risk prediction models (see, e.g., \citet{diamond1979analysis}, \citet{gail1989projecting}, \citet{tyrer2004breast}). After observing a test result, they are directed to use Bayes' theorem and the test's false positive and false negative rates to update the pre-test probability to a \textit{post-test} probability.\footnote{Updating is commonly implemented using online calculators or graphical tools such as Fagan's nomograms \citep{fagan1975nomogram}. For examples of online tools, see \href{https://jamaevidence.mhmedical.com/data/calculators/LR_nomogram.html}{Link 1} and \href{https://www.medcentral.com/calculators/stats-data/post-test-probability-from-pre-test-probability-sensitivity-and-specificity}{Link 2} \textit{(last accessed 08/25/2026)}. See \autoref{sect:nomogram} for examples of nomograms.}

When the test's error rates are uncertain, the literature calls for applying Bayes' theorem to every admissible pair of rates -- that is, full Bayesian updating -- yielding a set of post-test probabilities (see, e.g., \citealp{baron1994uncertainty}; \citealp{mossman2001intervals}; \citealp{bianchi2009incorporating}; \citealp{srinivasan2012propagation}; \citealp{manski-2021}). For some tests, the resulting set contains the original pre-test probability and additional values after \textit{either} result, so following the prescribed updating procedure induces dilation.

We first motivate the detection of dilation-inducing tests prior to their use in a common treatment decision problem. An informed maximin clinician weakly prefers not to administer such a test, whether they would follow the prescribed updating procedure or deviate from it and disregard the result. The preference is strict whenever testing carries \textit{any} positive cost. We then equivalently characterize conditions for a test to be dilation-inducing and highlight how a ubiquitous feature of test performance studies -- imperfect measurement of the true health status -- creates scope for these conditions to hold, even for tests that surpass previously used approval thresholds. Third, we develop a hypothesis test of whether a diagnostic test is dilation-inducing in this context. The hypothesis test is finite-sample valid over a broad and empirically relevant class of data-generating processes. Finally, we revisit a highly influential study that endorsed chest computed tomography (CT) scans as a primary tool for detecting COVID-19 early in the pandemic \citep{ai2020correlation} and contributed to their implementation \citep{de2020radiological}. The findings are consistent with the test being dilation-inducing, despite being recommended for use and exceeding an apparent-sensitivity benchmark used in FDA emergency-use authorization (EUA) guidance for COVID-19 tests.

Our motivating analysis studies a standard binary treatment-choice problem in which a clinician decides whether to administer a treatment that benefits the sick but harms the healthy. The clinician evaluates each decision by its worst-case expected utility over the post-test probabilities \citep{gilboa1989maxmin}. Maximin reasoning is consistent with some established clinical practices \citep{brody1981maximin,brody2013revisiting} and has been associated with the nonmaleficence ``\textit{do no harm}'' principle embedded in the Hippocratic tradition \citep{barrieu2006precautionary,hagmayer2017causal}.

For a dilation-inducing test, merely \textit{updating} as directed by existing guidance weakly lowers the clinician's value after either result, relative to disregarding it and acting on the pre-test probability. The loss is strict after at least one result. Testing costs are already sunk in this comparison: the loss comes from conditioning itself. This complements existing findings on the negative value of dilation-inducing information \citep{grunwald2005ignorance, shishkin2023ambiguous}. Clinical guidance may leave clinicians exposed to this loss -- it typically treats observed results as inputs into the updating procedure without explicitly offering the option of disregarding them. Nevertheless, whether they would use the result or disregard it, clinicians informed that a test is dilation-inducing weakly prefer \textit{not} to administer it, and strictly so whenever testing carries \textit{any} positive cost.

To enable detection, we provide an equivalent characterization of when a test is dilation-inducing. The condition is that the identified set of true positive and true negative rates used for updating contains one pair whose components sum to one and another whose components do not. This criterion is independent of the clinician's pre-test probability. We highlight that a ubiquitous feature of test performance studies -- measuring health status with an imperfect reference test -- creates scope for this condition to hold under standard assumptions. Moreover, in such studies, \textit{apparent performance}, performance measured against the reference test, is typically used as a convenient substitute for true performance in test approval and validation. A test can surpass such performance benchmarks while still being dilation-inducing.

\Cref{sect:inference} develops a hypothesis test of the null that a diagnostic test is dilation-inducing when its performance is evaluated against an imperfect reference test. The test is finite-sample valid over a broad and empirically relevant class of data distributions in which all four combinations of index and reference test results have positive probability. We exploit the exact multinomial distribution of the study's outcome counts and the method of \citet{berger1994p} to control the probability of falsely rejecting dilation at any sample size. In simulations, our test is less conservative and generally more powerful than alternatives based on projection \citep{romano2014practical} and subvector inference \citep{bugni2017inference} for moment inequality models, with substantial power gains in small samples.

Finally, we apply our method to study the use of chest CT scans for detecting COVID-19. We revisit the influential study of \citet{ai2020correlation}, which recommended CT as a primary detection tool in epidemic areas based on its estimated apparent sensitivity of $96.5\%$ relative to an RT-PCR reference test. This exceeded an estimated apparent-sensitivity benchmark used in FDA emergency-use authorization guidance for COVID-19 tests. The study also contributed to the adoption of chest CT screening at some Dutch hospitals early in the pandemic \citep{de2020radiological}. Nevertheless, our findings are consistent with the test being dilation-inducing. Because our method can be applied to performance-study data available before broader clinical use, it can help inform decisions about whether to implement a test. This is particularly relevant given subsequent work reporting limited diagnostic performance and recommending against routine CT screening, which had been discontinued in certain settings \citep{lee2022diagnostic}. Moreover, the method could help assess the performance evidence supporting the selective diagnostic uses of CT imaging retained in WHO guidance for symptomatic patients when virological testing is unavailable, delayed, or initially negative despite high clinical suspicion \citep{who2022imaging}.

This paper first contributes to the broad literature on dilation by showing that it can arise in practice from following established clinical guidance. Dilation has been studied as a theoretical phenomenon in statistics and imprecise probability theory \citep[e.g.,][]{good1974little, walley1991statistical,seidenfeld-wasserman-1993, herron1997divisive, gong2021judicious}, philosophy of probability and formal epistemology \citep[e.g.,][]{bradley2019imprecise,pedersen2014demystifying}, and economic theory. Economic applications include mechanism design \citep{bose2014mechansim}, persuasion \citep{beauchene2019ambiguous, cheng2023ambiguous, pahlke2024dynamic}, \citeauthor{aumann1976}'s (\citeyear{aumann1976}) agreement theorem \citep{zhang2018}, and combining forecasts \citep{levy2022}. \citet{manski2018credible} suggests the possibility of dilation in ecological inference for personalized patient care. More broadly, a recent significant body of experimental literature studies how individuals respond to ambiguous information \citep[e.g.,][]{hayashi2010,kellner2022reacting,shishkin2023ambiguous,epstein2024hard,kops2023testing,liang2023learning}. We complement this work by showing that dilation can arise in a real-world setting from standard procedures for interpreting diagnostic evidence.

Second, the paper contributes to the literature on diagnostic test performance by further highlighting the importance of considering true, rather than apparent, test performance when evaluating tests for approval. This connects to work on gold standard bias \citep{gart1966comparison2,staquet1981methodology,zhou2009statistical} and, more specifically, to work showing that true test performance measures are typically only partially identified under tenable assumptions \citep{thibodeau1981evaluating, emerson2018biomarker}. Finally, we provide regulators and researchers with tools for detecting dilation-inducing tests prior to their widespread clinical use.

\Cref{sect:setting} outlines the setting. \Cref{sect:motivating_model} motivates detecting dilation-inducing tests prior to their use. \Cref{sect:dilation_identification} characterizes conditions for dilation and shows that these conditions may arise from measuring test performance against an imperfect reference. \Cref{sect:inference} proposes a finite-sample valid hypothesis test for dilation. \Cref{sect:applications} applies the findings to revisit the use of CT chest scans for COVID-19 detection. Appendix \ref{app:proofs} collects all proofs.

\section{Setting}\label{sect:setting}

We consider a canonical prediction problem in clinical medicine where a clinician uses a test result to determine a patient's health status. This section first reviews the standard clinical practice for obtaining the underlying central parameters -- post-test probabilities. It then introduces the phenomenon of dilation and highlights that following this practice as instructed can lead to it.

Let $y=1$ denote the presence of the underlying illness of interest and $y=0$ its absence. Similarly, let $t=1$ denote a positive result and $t=0$ a negative result from a binary classifier intended to provide information about $y$. Typically, $t$ is a binary diagnostic test result, or a thresholded output of either a predictive model or a continuous diagnostic test. Following common nomenclature, we refer to $t$ as the \textit{index test} and to $y$ as the \textit{health status}.

Virtually any test can produce false positive or false negative results, so $t$ need not coincide with $y$. Let $Q(t,y)$ denote the joint distribution of $(t,y)$ in the clinical population from which the patient is drawn.  The probabilities of the test errors are governed by the following test performance parameters (\citealp{yerushalmy1947statistical}, \citealp{binney2021origin}):
\begin{align}
\theta_1 &:= Q(t=1 \mid y=1) \qquad \text{(sensitivity)}, \label{eq:def_{t'}ens}\\
\theta_0 &:= Q(t=0 \mid y=0) \qquad \text{(specificity)}. \label{eq:def_{t'}pec}
\end{align}

Let $\theta:=(\theta_0,\theta_1)$ and $\theta\in\Theta$, where $\Theta\subseteq[0,1]^2$ is possibly a singleton identified set obtained from a test performance study relevant to the clinical population. Regulatory approval or validation of $t$ typically requires such a study to demonstrate satisfactory test performance prior to its wider use.  For simplicity of exposition, we first take $\Theta$ as given by the performance study and return to how the set may be obtained in \Cref{sect:dilation_identification}. By treating the identified set as known, we also initially abstract from statistical imprecision and focus on dilation as an inherent joint property of the test and the performance study. In Sections \ref{sect:dilation_identification} and \ref{sect:inference}, we discuss detecting dilation from sample data.

\subsection{Standard Clinical Practice and Dilation}

Clinicians are typically instructed to approach determining the patient's health status as follows (examples are numerous; see, for instance, \citealp{mcneil1975primer} and \citealp{watson-et-al-2020}). First, they examine the patient and assess the \emph{pre-test probability} $\pi := Q(y=1) \in (0,1)$ that the patient has the underlying condition. This assessment is based on expert knowledge, prevalence rates, or risk prediction models (see, e.g., \citealp{diamond1979analysis}, \citealp{gail1989projecting}, \citealp{tyrer2004breast}).\footnote{Whether a test is dilation-inducing will not depend on the pre-test probability $\pi\in(0,1)$ assessed by the clinician. As is standard, we exclude $\pi\in\{0,1\}$, since the patient's health status is then already known with certainty and the test is unnecessary.} The clinician then observes a realization of the test $t$, and uses $\Theta$ from the relevant performance study to update $\pi$ via Bayes' theorem to the corresponding \textit{post-test} probabilities. Specifically, for any $\theta \in \Theta$ with $\theta_0+\theta_1\neq 1$, the post-test probabilities are:
\begin{align*}
	v_1(\theta;\pi) &:= \frac{\theta_1\pi}{\theta_1 \pi + (1-\theta_0)(1-\pi)},\\
	v_0(\theta;\pi) &:=\frac{(1-\theta_1)\pi}{\theta_0 (1-\pi) + (1-\theta_1)\pi},
\end{align*}
and $v_j(\theta;\pi)=\pi$ for test results $j \in \{0,1\}$ when $\theta_1+\theta_0=1$, which is equivalent to $t\independent y$; $v_j(\theta;\pi)$ represents $Q(y=1|t=j)$ given $(\theta,\pi)$. The condition $\theta_0+\theta_1\neq1$ ensures that the post-test probabilities are always well defined by Bayes' rule. When $\theta_0+\theta_1=1$, the test is uninformative and Bayesian updating after any realized result leaves $\pi$ unchanged.

Existing medical literature directs clinicians to do full Bayesian updating -- that is, derive the Bayesian update of $\pi$ for every possible value of $\theta$ -- when it is uncertain (see, e.g., \citealp{baron1994uncertainty}; \citealp{mossman2001intervals}; \citealp{bianchi2009incorporating}; \citealp{srinivasan2012propagation}; \citealp{manski-2021}). 
To this end, let the set of all post-test probabilities given $\Theta$ be
\begin{align}\label{eq:post_test_definition}
	V_j(\Theta;\pi) := \Big\{ v_j(\theta;\pi) : \theta \in \Theta \Big\} \qquad \text{for } j\in\{0,1\}.
\end{align}

Operationally, the post-test probabilities $v_j(\theta;\pi)$, and hence $V_j(\Theta;\pi)$, are typically obtained using an online calculator or a graphical tool such as a nomogram \citep{fagan1975nomogram}. See \autoref{sect:nomogram} for examples. Once $V_j(\Theta;\pi)$ is determined, it forms the basis for clinical decisions. \Cref{sect:motivating_model} provides an example.

\begin{remark}
Although the analysis can be extended to settings in which $y$ and $t$ have finite supports, we restrict attention to binary tests and binary health statuses to align with current research practice. Prominent statistical guidelines, such as \citet{fda2007statistical}, are formulated for binary $t$ and $y$. In practice, various tests, such as RT-PCR or biomarker tests, are reduced to binary outcomes by thresholding. 
\end{remark}

To build intuition for the definition of dilation, consider the post-test probability set $V_j(\Theta;\pi)$. When $\Theta$ is a singleton, $V_j(\Theta;\pi)$ is also a singleton. \autoref{fig:point_pv} illustrates this case when $\theta_1+\theta_0>1$, so that $t$ performs better than a coin toss. A positive test result yields a post-test probability above $\pi$, whereas a negative test result yields a post-test probability below $\pi$. 

If $\Theta$ is not a singleton, then $V_j(\Theta;\pi)$ is generally not a singleton. In that case, we say that the test $t$ provides \textit{ambiguous} information. Nevertheless, $t$ may still be informative, as illustrated in \autoref{fig:set_pv}. Upon observing $t=1$, all possible values of the post-test probability lie above $\pi$, whereas upon observing $t=0$, all possible values lie below $\pi$.

For some non-singleton $\Theta$, $\pi$ may be strictly contained in the set of possible post-test probabilities for \textit{every} test result $t$, as depicted in \autoref{fig:dilation}. In that case, conditioning on $t$ dilates the pre-test probability for any result, and thus \textit{only} introduces ambiguity regarding the patient's health status. This is the central idea behind the phenomenon known as \textbf{dilation}. This indicates that updating the pre-test probability according to standard clinical protocol may lead to dilation for some tests, which we formalize in the following definition.

\begin{figure}[ht!]
	\centering
	\begin{subfigure}[b]{0.32\textwidth}
		\centering
		\begin{tikzpicture}[line cap=round, line join=round, x=5cm, y=5.0cm, scale=0.75, dot/.style={circle,fill=black,minimum size=4pt,inner sep=0pt,  outer sep=-1pt}]
			\draw [->,color=black] (0,-0.025) -- (0,1.05) node [above] {\footnotesize $Q(y=1)$};
			\draw [->,color=black] (0.67,-0.025) -- (0.67,1.05) node [above] {\footnotesize $Q(y=1|t)$};
			\draw [-,color=black] (0.015,1) -- (-0.015,1) node [left] {$1$};
			\draw [-,color=black] (0.6815,1) -- (.655,1);
			\draw [-,color=black] (0.015,0) -- (-0.015,0) node [left] {$0$};
			\draw [-,color=black] (0.6815,0) -- (.655,0);
			\draw[black, fill=black, opacity=1] (-0.03,0.5025) rectangle (0.03,0.4975);
			\draw[preaction={fill, blue, opacity=1}, black,pattern=horizontal lines,pattern color=black] (0.64,0.725) rectangle (.7,0.73);
			\draw[preaction={fill, red, opacity=1}, black,pattern=vertical lines,pattern color=black] (0.64,0.275) rectangle (0.7,0.28);
			\draw [dotted,color=black!25] (-0.05,0.5) -- (.7,0.5);
			\draw[dashed,color=black!33] (0.0,1) -- (0.67,1);
			\draw[dashed,color=black!33] (0.0,0) -- (0.67,0);
			\draw [->,>=stealth,out=45,in=180,color=blue, thick] (0.1,0.5) to node [yshift=5, midway, above, blue] {$t=1$} (0.6,0.725);
			\draw [->,>=stealth,out=-45,in=180,color=red, thick] (0.1,0.5) to node [yshift=-5, midway, below, red] {$t=0$} (0.6,0.275);
			\node[left] at (-0.03,0.5) {$\pi$};
			\node[right] at (0.7,0.725) {$V_1$};
			\node[right] at (0.7,0.275) {$V_0$};
		\end{tikzpicture}
		\caption{Unambiguous info.}
		\label{fig:point_pv}
	\end{subfigure}
	\hfill
	\begin{subfigure}[b]{0.32\textwidth}
		\centering
		\begin{tikzpicture}[line cap=round, line join=round, x=5cm, y=5.0cm, scale=0.75, dot/.style={circle,fill=black,minimum size=4pt,inner sep=0pt,  outer sep=-1pt}]
			\draw [->,color=black] (0,-0.025) -- (0,1.05) node [above] {\footnotesize $Q(y=1)$};
			\draw [->,color=black] (0.67,-0.025) -- (0.67,1.05) node [above] {\footnotesize $Q(y=1|t)$};
			\draw [-,color=black] (0.015,1) -- (-0.015,1) node [left] {$1$};
			\draw [-,color=black] (0.6815,1) -- (.655,1);
			\draw [-,color=black] (0.015,0) -- (-0.015,0) node [left] {$0$};
			\draw [-,color=black] (0.6815,0) -- (.655,0);
			\draw[black, fill=black, opacity=1] (-0.03,0.5025) rectangle (0.03,0.4975);
			\draw[preaction={fill, blue, opacity=0.35}, rounded corners=2pt, pattern=horizontal lines,pattern color=black] (0.64,0.6) rectangle (0.7,0.85);
			\draw[preaction={fill, red, opacity=0.35}, rounded corners=2pt, pattern=north west lines,pattern color=black] (0.64,0.4) rectangle (0.7,0.15);
			\draw [dotted,color=black!25] (-0.05,0.5) -- (.7,0.5);
			\draw[dashed,color=black!33] (0.0,1) -- (0.67,1);
			\draw[dashed,color=black!33] (0.0,0) -- (0.67,0);
			\draw [->,>=stealth,out=45,in=180,color=blue, thick] (0.1,0.5) to node [yshift=5, midway, above, blue] {$t=1$} (0.6,0.725);
			\draw [->,>=stealth,out=-45,in=180,color=red, thick] (0.1,0.5) to node [yshift=-5, midway, below, red] {$t=0$} (0.6,0.275);
			\node[left] at (-0.03,0.5) {$\pi$};
			\node[right] at (0.7,0.725) {$V_1$};
			\node[right] at (0.7,0.275) {$V_0$};
		\end{tikzpicture}
		\caption{Ambiguous info.}
		\label{fig:set_pv}
	\end{subfigure}
	\hfill
	\begin{subfigure}[b]{0.32\textwidth}
		\centering
		\begin{tikzpicture}[line cap=round, line join=round, x=5cm, y=5.0cm, scale=0.75, dot/.style={circle,fill=black,minimum size=4pt,inner sep=0pt,  outer sep=-1pt}]
			\draw [->,color=black] (0,-0.025) -- (0,1.05) node [above] {\footnotesize $Q(y=1)$};
			\draw [->,color=black] (0.67,-0.025) -- (0.67,1.05) node [above] {\footnotesize $Q(y=1|t)$};
			\draw [-,color=black] (0.015,1) -- (-0.015,1) node [left] {$1$};
			\draw [-,color=black] (0.6815,1) -- (.655,1);
			\draw [-,color=black] (0.015,0) -- (-0.015,0) node [left] {$0$};
			\draw [-,color=black] (0.6815,0) -- (.655,0);
			\draw[black, fill=black, opacity=1] (-0.03,0.5025) rectangle (0.03,0.4975);
			\draw[preaction={fill, blue, opacity=0.35}, rounded corners=2pt, pattern=horizontal lines,pattern color=black] (0.64,0.4) rectangle (0.7,0.85);
			\draw[preaction={fill, red, opacity=0.35}, rounded corners=2pt, pattern=north west lines,pattern color=black] (0.64,0.6) rectangle (0.7,0.15);
			\draw [dotted,color=black!25] (-0.05,0.5) -- (.7,0.5);
			\draw[dashed,color=black!33] (0.0,1) -- (0.67,1);
			\draw[dashed,color=black!33] (0.0,0) -- (0.67,0);
			\draw [->,>=stealth,out=45,in=180,color=blue, thick] (0.1,0.5) to node [yshift=5, midway, above, blue] {$t=1$} (0.6,0.725);
			\draw [->,>=stealth,out=-45,in=180,color=red, thick] (0.1,0.5) to node [yshift=-5, midway, below, red] {$t=0$} (0.6,0.275);
			\node[left] at (-0.03,0.5) {$\pi$};
			\node[right] at (0.7,0.725) {$V_1$};
			\node[right] at (0.7,0.275) {$V_0$};
		\end{tikzpicture}
		\caption{Dilation}
		\label{fig:dilation}
	\end{subfigure}
	\caption{Updating pre-test to post-test probabilities. $V_j$ is a shorthand for $V_j(\Theta;\pi)$ where $j\in\{0,1\}$.}
    \caption*{\justifying{The left panel depicts a singleton $\Theta$ with $\theta_1+\theta_0> 1$. In the middle panel, $\Theta$ is non-singleton, but $t$ is informative. The right panel presents dilation.}}
	\label{fig:pv_illustration}
\end{figure}

\begin{definition} \label{def:dilation}
	 $t$ is \emph{dilation-inducing} if for all $\pi\in(0,1)$, $\{\pi\} \subsetneq V_1(\Theta;\pi) \text{ and } \{\pi\} \subsetneq V_0(\Theta;\pi)$.
\end{definition}

In other words, $t$ \textit{induces dilation} if any nondegenerate pre-test probability $\pi$ is strictly contained in the set of possible post-test probabilities, regardless of the test outcome. For simplicity, we restrict attention to a point-valued pre-test probability in the main analysis. The original definition of dilation in \citet{seidenfeld-wasserman-1993} allows for $\pi$ to be partially identified. Adapting \autoref{def:dilation} and the main results to accommodate this case is straightforward (see \autoref{sect:pre_test_ambiguity}).

\begin{remark}
    Whether dilation is possible critically depends on the updating rule. Updating via the proxy or the contraction rules \citep{gul2021evaluating, tang2022theory} would preclude dilation. The full Bayesian rule \citep{pacheco2002rule, epstein2003recursive} in \eqref{eq:post_test_definition} underlies the clinical decision-making practice and is a rule conducive to dilation. 
\end{remark}

\section{Why Does Dilation Matter?}\label{sect:motivating_model}

To motivate detecting dilation-inducing tests prior to their use, we study them in a common treatment decision setting. Consider a clinician deciding whether to treat or not treat a patient, denoted
by $d\in D:=\{T,NT\}$. The patient's health outcome, $h_y^d\in\mathbb{R}$, depends on the decision $d$ and the patient's true health status $y\in\{0,1\}$. Assume treatment is beneficial when the patient is sick, detrimental when the
patient is healthy, and being healthy is always better than being sick so that
\[
    h^{NT}_0 > h^T_0 > h^T_1 > h^{NT}_1.
\]
The clinician evaluates health outcomes using a strictly increasing utility function $u:\mathbb{R}\to\mathbb{R}$. This allows for general risk attitudes while preserving a preference for better health outcomes.\footnote{The decision problem formulation is standard. For related examples see \citet{phelps1988focusing}, \citet{berger2013treatment}, and \citet{manski2013diagnostic}.}

In line with clinical guidance, the clinician first assesses a pre-test probability $\pi\in(0,1)$. Before choosing $d$, they may administer a test and update $\pi$ as described in the previous section. Let $\tau=1$ indicate that the test is administered, and $\tau=0$ otherwise. Let $c\geq0$ denote the cost of testing in utility terms. This captures the direct costs that virtually any test may impose by being invasive, psychologically taxing, financially burdensome, or harmful through delays in treatment. We allow $c=0$ to highlight that dilation may generate losses beyond the cost of testing. The expected utility of treatment decision $d$, given illness probability $p\in[0,1]$ and test status $\tau\in\{0,1\}$, is
\[
    U(d,p,\tau)
    :=
    p u(h_1^d)+(1-p)u(h_0^d)-c\tau.
\]

Established clinical practices suggest that clinicians may act in line with maximin reasoning under ambiguity \citep{brody1981maximin,brody2013revisiting}. Maximin reasoning has also been associated with precautionary decision-making and, in medical examples, with the nonmaleficence ``\textit{do no harm}'' principle embedded in the Hippocratic tradition \citep{barrieu2006precautionary,hagmayer2017causal}. Accordingly, we model the clinician as a maximin expected utility agent \citep{gilboa1989maxmin}.

Consider an interim problem commonly contemplated in clinical decision-making. The test has already been administered, and the clinician has observed a result. Clinical guidelines typically treat observed test results as inputs into the updating procedure, without formally offering the option of disregarding them. We ask whether, if given the choice, the clinician would prefer to follow the standard protocol or instead disregard the result. For a realized result $t=j\in\{0,1\}$, a maximin clinician following the standard updating protocol chooses $d$ that attains the optimal robust value
\[
    W_j
    :=
    \max_{d\in D}\inf_{p\in V_j(\Theta;\pi)}U(d,p,1).
\]

We compare this with a hypothetical benchmark in which the clinician disregards the observed result and instead acts on the original pre-test probability. The corresponding optimal value is
\[
    W^\pi
    :=
    \max_{d\in D}U(d,\pi,1).
\]

Let $\Delta_j:=W^\pi-W_j$ denote the loss from updating on result $j\in\{0,1\}$. The next proposition shows that, for a dilation-inducing test, updating on \textit{any} result never yields a higher value than disregarding it. For at least one result, disregarding the result yields a strictly higher value. Under a mild interiority condition, updating on either result makes a maximin clinician \textit{strictly} worse off than the counterfactual benchmark in which the result is ignored. The condition only rules out the knife-edge case in which $\pi$ lies on the boundary of the post-test probability sets for both results. 

\begin{proposition}\label{prop:post_test_updating_dilation}
    Suppose that $t$ is dilation-inducing. Then for any $c\geq0$:
    \begin{enumerate}
        \item $\Delta_j\geq 0$ for each $j\in\{0,1\}$, and $\Delta_{j'}>0$ for some $j'\in\{0,1\}$.
        \item $\Delta_j>0$ for each $j\in\{0,1\}$ if
        $\pi\in\inter\big(V_{j'}(\Theta;\pi)\big)$ for some $j'\in\{0,1\}$.
    \end{enumerate}
\end{proposition}

Note that at this interim stage the cost is sunk, i.e.\ both continuation values include the same testing cost, so each $\Delta_j$ is independent of $c$. The proposition therefore isolates the loss generated by conditioning on a result from a dilation-inducing test as prescribed by the standard protocol. This complements \citet[Proposition 1]{shishkin2023ambiguous} in a clinical setting. They demonstrate strictly negative information premia from dilation-inducing signals for maximin agents using full Bayesian updating under a more specialized signal structure. It also extends the complete-ignorance case studied by \citet{grunwald2005ignorance}, corresponding to $\Theta=[0,1]^2$, in which ignoring the signal is maximin-optimal.

Additional information making an agent worse off may seem like an undesirable feature of the decision problem, since the information can in principle be ignored. Indeed, \citet{shishkin2023ambiguous} provide evidence from an Ellsberg urn experiment consistent with such behavior. We highlight, however, that this feature can arise from following clinical guidance. As mentioned, observed test results are typically treated as inputs into a Bayesian updating procedure, without formally offering the option of not updating, potentially leaving clinicians susceptible to the loss from conditioning.

The interim result raises a natural question: would informing a protocol-abiding
clinician that a test is dilation-inducing deter testing? Suppose the
clinician is informed \textit{ex ante} that the test is dilation-inducing.
Assume they anticipate that, after observing $t=j$, they will follow
standard protocol and solve the continuation maximin problem defining
$W_j$. This postulates that the clinician correctly anticipates their continuation choices, as in \citet{strotz1955} (see also \citealp{siniscalchi2011}). Administering the test therefore generates the contingent
continuation values $(W_0,W_1)$. If the clinician does not administer the test, the optimal value is
\[
    W^N
    :=
    \max_{d\in D}U(d,\pi,0).
\]
Note that $W^N = W^\pi+c = W_j+\Delta_j+c$ for $j\in\{0,1\}$. \textit{Ex-ante} information thus enables this clinician to avoid two sources of loss by not administering the test: ($1$) cost $c$ from administering the test and ($2$) $\Delta_j$ from applying the protocol. Since \(\Delta_j \geq 0\) for every result and \(\Delta_j>0\) for some \(j\in\{0,1\}\), not administering the test yields a weakly higher value for every possible test result and a strictly higher value for at least one, even if the test is \emph{costless}.

\begin{corollary}\label{prop:ex_ante_dilation1}
    Suppose that the clinician anticipates following the standard updating protocol after any result. If $t$ is dilation-inducing, then, for any \(c\geq 0\), administering the test lowers the clinician's value for at least one possible result and raises it for none.
\end{corollary}

It is also immediate that a clinician who anticipates disregarding any test result that generates a loss from updating would strictly prefer not to administer the test whenever $c>0$, while administering is without loss \textit{only if} the test is costless $c=0$. By \Cref{prop:post_test_updating_dilation} and the definition of $\Delta_j$, administering the test would generate continuation values $(W^\pi,W^\pi)$. Since $W^N=W^\pi+c$, the result follows directly.

\begin{corollary}\label{prop:ex_ante_dilation2}Suppose that \(t\) is dilation-inducing and that the clinician anticipates deviating from the protocol -- i.e.\ ignoring the test result -- if and only if \(\Delta_j>0\). Then, relative to not administering the test, administering it lowers the clinician's value for \emph{every} possible result when \(c>0\) and leaves it unchanged for every possible result only when \(c=0\).
\end{corollary}

\section{When is a Test Dilation-Inducing?}\label{sect:dilation_identification}

We now equivalently characterize when a test is dilation-inducing. The characterization makes clear when dilation is a property of the index test and the underlying performance study, and forms the basis for detecting such tests from study data.

\begin{proposition}\label{prop:dilation_characterization}
The index test is dilation-inducing if and only if there exist $\theta, \theta' \in \Theta$ such that $\theta_0+\theta_1=1$ and $\theta_0'+\theta_1'\neq 1$.
\end{proposition}

The condition $\theta_0+\theta_1=1$ means that the performance study is
consistent with an uninformative index test, for which
$t\independent y$. This ensures that
$\pi\in V_j(\Theta;\pi)$ for either result $j$, and subsumes the
necessary condition for dilation of \citet[Theorem 2.1]{seidenfeld-wasserman-1993}.
The existence of some $\theta'\in\Theta$ such that
$\theta'_0+\theta'_1\neq1$ means that the study is also consistent with
an index test that provides some information about $y$. Each post-test probability set therefore also
contains a value different from $\pi$. Thus, dilation arises precisely
when the performance study cannot distinguish an uninformative index test
from an informative one.

Accordingly, $\Theta$ must be non-singleton for $t$ to be dilation-inducing. Partial identification of $\theta$ is therefore \textit{necessary} (but not sufficient) for dilation to be an inherent joint property of the index test and the underlying performance study. \Cref{prop:dilation_characterization} forms the basis for inference about whether the index test is inherently dilation-inducing as evaluated by the performance study. If the population distribution determining $\Theta$ were known, these conditions could be verified directly. More commonly, only sample data are available. The characterization can then be used to formulate a statistical test of whether the property holds. 

We now highlight that a ubiquitous characteristic of test performance studies frequently leads to partial identification of $\theta$ under tenable assumptions, making it possible for $t$ to be dilation-inducing. Importantly, this can occur with tests that surpass benchmarks previously used for test approval or recommendation. The following section then uses \Cref{prop:dilation_characterization} to develop a corresponding finite-sample valid hypothesis test.

\begin{remark}
In practice, clinicians may approximate the population-level  updating procedure using a sample-based set $\widehat\Theta_n$, such as a confidence set for $\theta$, by forming $V_j(\widehat\Theta_n;\pi)$ as a proxy for $V_j(\Theta;\pi)$ (e.g. \citealp{baron1994uncertainty, mossman2001intervals}). Statistical imprecision may then cause updating based on $\widehat\Theta_n$ to produce dilation, even when $t$ is not dilation-inducing, including when $\theta$ is point identified. Whether dilation arises from a particular sample-based set is distinct from whether $t$ is dilation-inducing, and the two may differ in how they can be eliminated. Dilation generated by statistical imprecision may disappear as the performance study sample size increases. By contrast, if $t$ is dilation-inducing, dilation is an inherent population-level property of the index test and the performance study. Eliminating it requires additional identifying information, which may not be plausibly available.

\Cref{prop:dilation_characterization} can reveal both types of dilation: applied to the identified set $\Theta$, it determines whether $t$ is dilation-inducing; applied to $\widehat\Theta_n$, it determines whether updating based on the specific sample set produces dilation.
\end{remark}

\subsection{Test Performance Studies and Imperfect References}

Thus far, we have taken $\Theta$ as given. We now turn to how it is obtained under common assumptions and widespread study designs. 
A pervasive feature of test performance studies is that the true health status $y$ cannot be observed directly and is instead measured using an imperfect \textit{reference test}. Under tenable assumptions, $\theta$ then generally remains partially identified \citep{thibodeau1981evaluating, emerson2018biomarker}, creating scope for $t$ to be dilation-inducing.\footnote{$\theta$ may be partially identified for other reasons, as well. Another source of partial identification is missing test data for either $t$ or $r$. This typically arises under study designs with test-result-based sampling, where one test is administered conditional on the result of the other. Such designs are generally avoided, so we abstract from them here \citep{kohn2022studies}. Nevertheless, \Cref{prop:dilation_characterization} can serve as a basis for detecting dilation-inducing tests regardless of the source of partial identification.} 

Let $r\in\{0,1\}$ denote the reference test, where $r=1$ and $r=0$ indicate positive and negative reference-test results, respectively. Let $P(t,r,y)$ denote the population distribution in the performance study, where $y$ is latent. We distinguish this distribution from the clinical population distribution $Q(t,y)$ relevant for the post-test probabilities of interest. This distinction makes explicit the standard practice of extrapolating $\theta$ from the performance study to compute post-test probabilities in the clinical population.\footnote{\citet{mulherin-miller-2002} and \citet{willis-2008} discuss how the design of performance studies can improve generalizability and provide guidance for assessing whether performance study measures extrapolate to clinical populations of interest.} We maintain the assumption underlying this practice -- the performance of $t$ does not vary across the two populations, or $\theta_j:=Q(t=j| y=j)=P(t=j| y=j)$ for $j\in\{0,1\}$. The two populations may differ in other respects, such as disease prevalence $Q(y=1)\neq P(y=1)$.

Study data are collected by randomly sampling and testing each individual with both the reference and index tests. This reveals $(t,r)$, so the data identify the joint distribution $P(t,r)$. The reference test $r$ has its own performance measures $s_j:=P(r=j| y=j)$ for $j\in\{0,1\}$, which we denote by $s:=(s_0,s_1)$. The reference $r$ should ideally be the best available test for $y$. Nevertheless, it is almost always imperfect so $s_j<1$ for at least some $j\in\{0,1\}$. Hence, $P(r=y)<1$, and the observed distribution $P(t,r)$ does not identify $\theta$ without further assumptions \citep{zhou2009statistical}. The observed distribution does identify \textit{apparent} sensitivity and specificity $\tilde\theta:=(\tilde\theta_0,\tilde\theta_1)$, defined as $\tilde\theta_j:=P(t=j| r=j)$ for $j\in\{0,1\}$.

\begin{remark}
\textit{Apparent} performance $\tilde\theta$ is routinely used for test approval or validation as an operationally convenient substitute for $\theta$, even though it generally differs from \textit{true} test performance when $r$ is imperfect, so that $\tilde\theta\neq\theta$. For this reason, $\tilde\theta$ should \textit{not} be used to compute the post-test probabilities needed for clinical decision-making \citep{fda2007statistical}. Moreover, this practice can lead to the case in which $t$ passes standard performance benchmarks while still being dilation-inducing. See the discussion in \Cref{subsect:example_dil}.
\end{remark}

A sizeable literature studies identification of the true test performance $\theta$ when $r$ is imperfect. Early work point identifies $\theta$ by assuming exact knowledge of $s$, motivated by the fact that $r$ is typically a well-established test, together with conditional independence of $t$ and $r$, i.e.\ $t\independent r|y$ (\citealp{buck1966comparison1}, \citealp{staquet1981methodology}). Several authors, however, argue that conditional independence is implausible in practice (e.g.\ \citealp{vacek1985effect}, \citealp{valenstein1990evaluating}, \citealp{hui1998evaluation}). \citet{thibodeau1981evaluating} and \citet{emerson2018biomarker} therefore allow $t\notindependent r|y$, which generally partially identifies $\theta$ when $s$ is known. \citet{obradovic-2022} derives the sharp identified set under the same maintained assumptions and also allows $s$ to be known only approximately. We inherit these common assumptions from the imperfect-reference-test literature.

\begin{assumption}{(Random Sampling)}\label{random_sampling_assumption}
The study sample is a sequence of i.i.d. random vectors $W_i=(t_i,r_i)$, where each $W_i$ follows a categorical distribution $P(t,r)$ for $(t,r)\in\{0,1\}^2$ and $i=1,\hdots,n$.
\end{assumption}

\begin{assumption}{(Reference Performance)}\label{assumption:reference_performance}
Sensitivity and specificity of the reference test are known and satisfy $s_1+s_0 > 1$.
\end{assumption}

\autoref{assumption:reference_performance} is particularly appealing when $s$ is identified from a separate study using a perfect reference test that is unsuitable for routine use as a comparator, such as post-mortem pathology for Alzheimer's disease \citep{suemoto2023autopsy}. It may also be justified by the physical or analytical properties of $r$; see \citet[Section 2.1]{obradovic-2022} for further discussion of both cases. The assumption also requires $s_1+s_0>1$, ruling out an uninformative reference test with $r\independent y$. Beyond this, $s_1>1-s_0$ is without loss of generality, since if $s_1<1-s_0$ one can relabel the reference result as $r^*=1-r$.

\begin{remark}
Knowledge of $s$, while commonly assumed in this literature, may be restrictive in some settings. Appendix \ref{app:extensions} therefore allows $s$ to be only approximately known, provides a tool for sensitivity analysis, and explains how the analysis can proceed without imposing explicit assumptions on $s$, beyond $s_0+s_1>1$.

\end{remark}

\begin{assumption}{(Bounded Prevalence)}\label{assumption:prevalence}
Population prevalence satisfies $P(y=1) \in (0,1)$.
\end{assumption}

Assumption \ref{assumption:prevalence} is implicit in any test performance study identifying $\theta$. Without it, one of the performance measures is not properly defined. Assumptions \ref{assumption:reference_performance} and \ref{assumption:prevalence} jointly point identify the study population prevalence since $P(y=1) = \frac{P(r=1)+s_0-1}{s_1+s_0-1}$. 
We henceforth use $P_s$ to denote identified probabilities derived from observable distributions given $s$, such as $P_s(y=1)$. Directly observed probabilities, such as $P(t,r)$, carry no subscript.

 Under Assumptions \ref{random_sampling_assumption}, \ref{assumption:reference_performance} and \ref{assumption:prevalence}, \citet[Proposition 1]{obradovic-2022} shows that the sharp identified set $\Theta_P(s)$ for $\theta$ given $P(t,r)$ and $s$ is
\begin{equation} \label{eq:H_s}
    \Theta_P(s) := \left\{
\begin{array}{cl}
   \vcenter{\hbox{\((\theta_0, \theta_1) \in [0,1]^2 :\)}} & \begin{aligned}
        &\theta_1 \in [\theta^L_1(s),\theta^U_1(s)] \text{ and}\\
        &\theta_0 = \frac{\theta_1 P_s(y=1)-P(t=1)}{P_s(y=0)}+1
    \end{aligned}
\end{array}
\right\},
\end{equation}
where
\begin{align}
\nonumber
    \theta_1^L(s) := \frac{1}{P_s(y=1)} 
    \bigg[
    &\max\left\{0,P(t=1, r=0)-s_0 P_s(y=0)\right\} 
    + \\
   &\max\left\{0,P(t=1, r=1)-(1-s_0)P_s(y=0)\right\}
    \bigg] \label{eq:theta1L}
\end{align}
and
\begin{align}
 \nonumber
 \theta_1^U(s) :=
  \frac{1}{P_s(y=1)} 
    \bigg[
    & \min\left\{P(t=1, r=0),(1-s_1)P_s(y=1)\right\} + \\
    & \min\left\{P(t=1, r=1),s_1P_s(y=1)\right\}
    \bigg]. \label{eq:theta1U}
\end{align}

The expression shows that $\theta$ may commonly be only partially identified, making $t$ conducive to being dilation-inducing. As we show in \autoref{sect:inference}, this occurs for a large and practically relevant class of $P(t,r)$ whenever $r$ is imperfect.

\subsubsection{Common Performance Benchmarks and Dilation-Inducing Tests}\label{subsect:example_dil}

Validation, recommendation, and authorization of diagnostic tests commonly rely on minimum performance measured relative to a reference test. When the reference is imperfect, these criteria concern \textit{apparent} performance $\tilde\theta$, whereas the post-test probabilities of interest depend on true performance $\theta$. For example, guidance from the World Health Organization (WHO) and the European Centre for Disease Prevention and Control (ECDC) used minimum apparent sensitivity and specificity of $80\%$ and $97\%$, respectively, relative to RT-PCR for COVID-19 rapid antigen tests \citep{world2020antigen, ecdc2021options}. Similar benchmarks have been used for rapid influenza diagnostic tests \citep{green2018rapid}. FDA Emergency Use Authorization (EUA) templates  for COVID-19 antigen tests likewise included an apparent-sensitivity benchmark of $80\%$ relative to RT-PCR.\footnote{See: \href{https://www.fda.gov/media/137907/download}{Link 1} and \href{https://www.fda.gov/media/156146/download?attachment}{Link 2} \textit{(last accessed 08/25/2026)}.} High apparent sensitivity has also been invoked to support recommendations for use even when apparent specificity is low \citep{ai2020correlation}. These criteria, while common, need not rule out dilation.

Consider the population distribution $P(t,r)$ in \autoref{tab:policy_example} first. It yields apparent specificity $\tilde{\theta}_0=38.8\%/40\%=97\%$ and apparent sensitivity $\tilde{\theta}_1=48\%/60\%=80\%$, meeting the WHO, ECDC and FDA benchmarks. Suppose that the reference test has perfect specificity, $s_0=1$, and sensitivity $s_1=0.61$. \Cref{fig:example1} shows that the identified set $\Theta_P(s)$ from \eqref{eq:H_s} is non-singleton and contains the point $(0.508,0.492)$ for which $\theta_1+\theta_0=1$. By \Cref{prop:dilation_characterization}, the test is dilation-inducing. Thus, a test can simultaneously be dilation-inducing and satisfy common performance benchmarks.

\begin{table}[ht!]
	\centering
	\caption{Two potentially worrisome examples}
	\label{tab:worrisome_examples}
	
	\begin{subtable}[t]{0.4\textwidth}
		\centering
		\begin{tabular}{@{}cccc@{}}
			\toprule
			$P(t\downarrow,r\rightarrow)$ & $r=0$ & $r=1$ & $P(t)$ \\ 
			\midrule
			$t=0$ & $38.8\%$ & $12.0\%$ & $50.8\%$\\
			$t=1$ & $1.2\%$ & $48.0\%$ & $49.2\%$\\ 
			\midrule
			$P(r)$ & $40\%$ & $60\%$ & \\
			\bottomrule
		\end{tabular}
		\caption{Example 1}
		\label{tab:policy_example}
	\end{subtable}
	\hspace{0.1em}
	\begin{subtable}[t]{0.42\textwidth}
		\centering
		\begin{tabular}{@{}cccc@{}}
			\toprule
			$P(t\downarrow,r\rightarrow)$ & $r=0$ & $r=1$ & $P(t)$\\
			\midrule
			$t=0$ & $10.4\%$ & $2.1\%$ & $12.4\%$\\
			$t=1$ & $30.4\%$ & $57.2\%$ & $87.5\%$\\
			\midrule
			$P(r)$ & $40.7\%$ & $59.3\%$ & \\
			\bottomrule
		\end{tabular}
		\caption{Example 2}
		\label{tab:ai_dgp}
	\end{subtable}
\end{table}

\pgfplotsset{
    thetaaxis/.style={
        width=0.43\textwidth,
        height=0.43\textwidth,
        xmin=0, xmax=1,
        ymin=0, ymax=1,
        xtick={0,0.2,0.4,0.6,0.8,1},
        ytick={0,0.2,0.4,0.6,0.8,1},
        tick label style={font=\footnotesize},
        grid=major,
        grid style={gray!20},
        axis line style={gray!60},
        tick style={gray!60},
        xlabel={$\theta_0$},
        ylabel={$\theta_1$},
        ylabel style={
            rotate=-90
        },
        %legend cell align={left},
    }
}

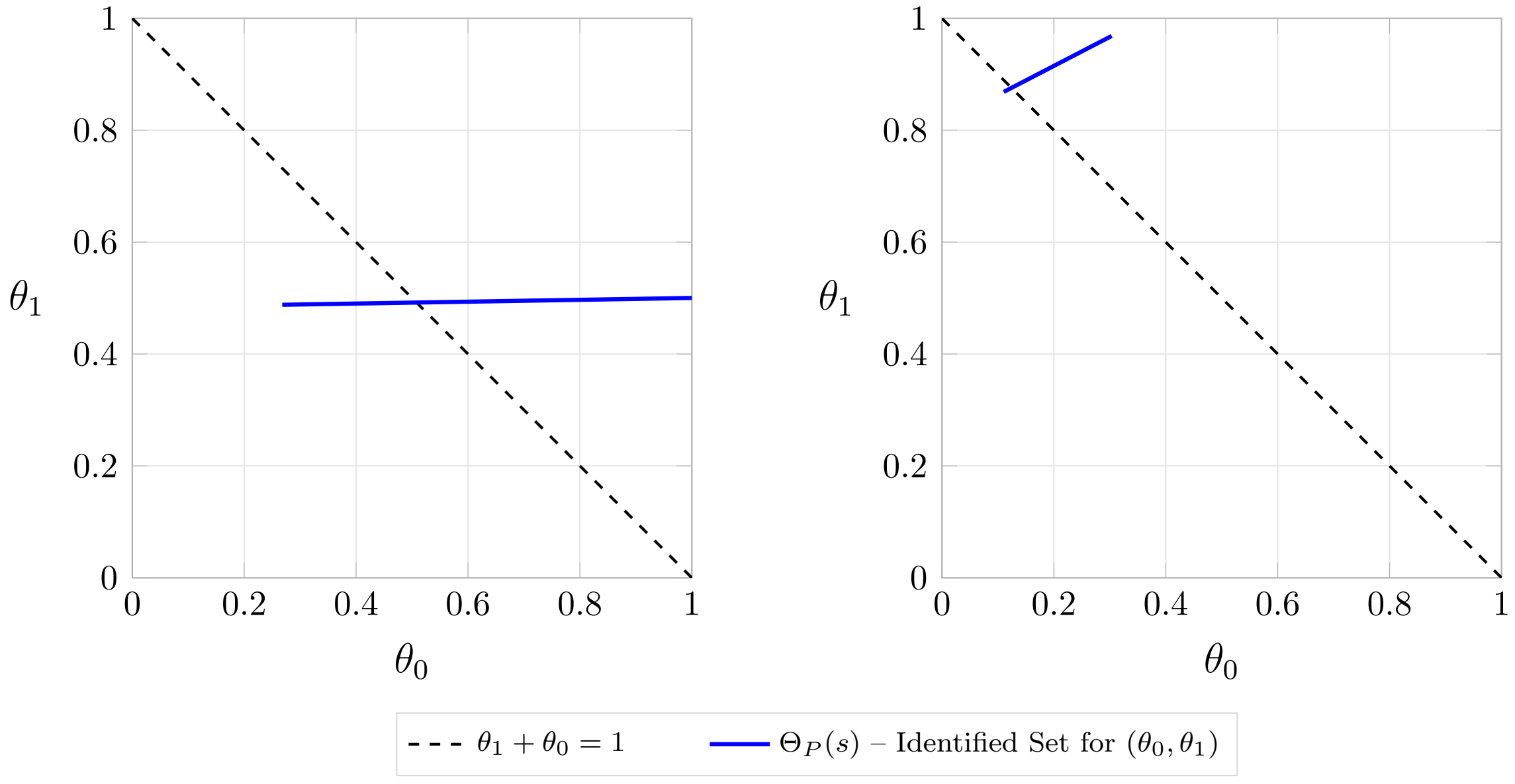
\begin{figure}[htbp]
\centering

\begin{tikzpicture}
% Subfigure counters, labels, and hyperlink anchors
\phantomsubcaption\label{fig:example1}
\phantomsubcaption\label{fig:example2}

\begin{groupplot}[
    group style={
        group size=2 by 1,
        horizontal sep=2.5cm
    },
    thetaaxis
]

% ---------- Panel (a) ----------
\nextgroupplot[
    legend to name=examplelegend,
    legend columns=2,
    legend style={
        draw=gray!30,
        fill=white,
        font=\scriptsize,
        inner sep=3pt,
        /tikz/every even column/.append style={column sep=0.75cm}
    }
]

\addplot[
    black,
    dashed,
    thick,
    domain=0:1,
    samples=2
] {1-x};
\addlegendentry{$\theta_1+\theta_0=1$}

\addplot[
    blue,
    very thick
]
coordinates {
    (0.268,0.488)
    (1.000,0.5002)
};;
\addlegendentry{$\Theta_P(s)$ -- Identified Set for $(\theta_0,\theta_1)$}

% ---------- Panel (b) ----------
\nextgroupplot

\addplot[
    black,
    dashed,
    thick,
    domain=0:1,
    samples=2
] {1-x};

\addplot[
    blue,
    very thick
]
coordinates {
    (0.110397946,0.868552413)
    (0.303273427,0.968552413)
};

\end{groupplot}

% First subfigure counter + label

\node[anchor=north]
    at ([yshift=-22mm]group c1r1.south)
    {(a) $s=(1,0.61)$};

\node[anchor=north]
    at ([yshift=-22mm]group c2r1.south)
    {(b) $s=(1,0.9)$};

    % Common legend
\node[anchor=north]
    at ([yshift=-12mm]$(group c1r1.south)!0.5!(group c2r1.south)$)
    {\pgfplotslegendfromname{examplelegend}};

\end{tikzpicture}

\caption{Identified sets in the examples}
\label{fig:examples}

\end{figure}

Example 1 relies on a reference test with relatively low sensitivity, although estimates in this range have been reported for some RT-PCR assays.\footnote{\citet{alcoba2020sensitivity} use repeated testing of the same samples to estimate sensitivity as low as $60.2\%$ for some COVID-19 RT-PCR assays.} However, the phenomenon of dilation-inducing tests which surpass established performance benchmarks does \textit{not} require an imprecise reference test. A complementary example demonstrates this.

Consider now the population distribution in \Cref{tab:ai_dgp}. Its apparent sensitivity is $\tilde{\theta}_1=57.2\%/59.3\%=96.5\%$, substantially surpassing the FDA EUA benchmark, although its apparent specificity is only $\tilde\theta_0 =10.4\%/40.7\%= 25.6\%$. Suppose that the reference test has perfect specificity, $s_0=1$, and sensitivity $s_1=0.9$. \Cref{fig:example2} and \Cref{prop:dilation_characterization} show that the index test is dilation-inducing while surpassing the FDA benchmark, despite the high accuracy of the reference test.

The low apparent specificity may make the example seem artificial. However, the population distribution in the example is set equal to the empirical distribution reported by the highly influential study of \citet{ai2020correlation}, which recommended the index test for use based on its high apparent sensitivity despite its low apparent specificity. In \Cref{sect:applications}, we return to the underlying sample and show that, under the same reference performance, the hypothesis that the index test is dilation-inducing cannot be rejected.

\section{Statistical Inference}\label{sect:inference}

We now turn to detecting dilation-inducing tests from performance study data. Regulatory bodies approving diagnostic tests, such as the FDA, typically require methods that account for statistical imprecision. We therefore construct a hypothesis test of whether $t$ is dilation-inducing, focusing on the ubiquitous setting of a performance study with an imperfect reference test. The test is finite-sample valid over a broad and empirically relevant class of data distributions.

\autoref{prop:dilation_characterization} forms the basis for the test. Applied to $\Theta_P(s)$, the index test is dilation-inducing if and only if there exist $\theta,\theta'\in\Theta_P(s)$ such that $\theta_1+\theta_0=1$ and $\theta_1'+\theta_0'\ne1$. Because $\Theta_P(s)$ is either a singleton or a line segment with strictly positive and finite slope, it can intersect the antidiagonal $\{\theta\in[0,1]^2:\theta_1+\theta_0=1\}$ at most once. Hence, $t$ is dilation-inducing if and only if $\Theta_P(s)$ contains a point on the antidiagonal and is not a singleton. We next show that the latter condition holds for a broad and practically relevant class of data-generating processes (DGPs) whenever $r$ is imperfect. The class is defined by the following assumption.

\begin{assumption}{(Full Support)}\label{assumption:everything_goes}
For every $(j,k)\in\{0,1\}^2$, $P(t=j,r=k)>0$.
\end{assumption}

\autoref{assumption:everything_goes} requires every combination of index- and reference-test results to occur with positive probability. This is a mild condition in conventional performance studies where each participant receives both tests. Full support is not required to identify $\Theta_P(s)$ or for the general characterization of dilation. We impose it here only to simplify the characterization underlying the statistical test.

\begin{theorem}\label{th:simpler_inference}
Maintain Assumptions \ref{random_sampling_assumption}, \ref{assumption:reference_performance},
\ref{assumption:prevalence}, and \ref{assumption:everything_goes}, and let
$\Theta_P(s)$ be the resulting identified set in \eqref{eq:H_s}. Then $t$ is
dilation-inducing if and only if (i) $s\ne(1,1)$ and (ii) there exists
$\theta\in\Theta_P(s)$ such that $\theta_1+\theta_0=1$.
\end{theorem}

Thus, when the reference test is imperfect, testing the null that $t$ is dilation-inducing is equivalent to testing $H_0: \exists\theta\in\Theta_P(s)
\text{ s.t. }\theta_0+\theta_1=1$ against $H_1: \theta_0+\theta_1\ne1
\text{ for every }\theta\in\Theta_P(s)$. General-purpose methods for inference in moment inequality models, such as \citet{romano2014practical} and \citet{bugni2017inference}, could deliver uniformly valid asymptotic inference. We instead exploit the structure of the problem to construct a finite-sample valid test. Let
\[
p_{jk}:=P(t=j,r=k),
\quad
p_{j\cdot}:=P(t=j),
\quad
p_{\cdot k}:=P(r=k),
\; \text{ and } \;
\Vec p:=(p_{00},p_{01},p_{10},p_{11}).
\]
Recall that, under Assumptions \ref{assumption:reference_performance} and
\ref{assumption:prevalence}, $P_s(y=1)
=
\frac{p_{\cdot1}+s_0-1}{s_0+s_1-1}.$ Define two observable gaps:
\begin{align*}
g_L(\Vec p;s)
&:=
p_{11}
-P_s(y=1)(p_{1\cdot}+s_1-1),\\
g_U(\Vec p;s)
&:=
p_{01}
-P_s(y=1)(s_1-p_{1\cdot}).
\end{align*}

The next proposition translates the identified-set condition in
\Cref{th:simpler_inference} into restrictions on the observable distribution.

\begin{proposition}
\label{prop:primitive-dilation-null}
Maintain Assumptions \ref{random_sampling_assumption}, \ref{assumption:reference_performance},
\ref{assumption:prevalence}, and \ref{assumption:everything_goes}, and suppose
$s\ne(1,1)$. The line containing $\Theta_P(s)$ intersects the
antidiagonal at the unique point $\theta^\star(P)=(1-p_{1\cdot},p_{1\cdot}).$
Moreover,
\[
\theta^\star(P)\in\Theta_P(s)
\quad\Longleftrightarrow\quad
g_L(\Vec p;s)\ge0
\quad\text{and}\quad
g_U(\Vec p;s)\ge0.
\]
Consequently, $t$ is dilation-inducing if and only if both gap inequalities
hold.
\end{proposition}

Denote by $\Delta_3$ the three-dimensional simplex. Let
$\mathcal P_s\subseteq\Delta_3$ be the set of observable distributions of
$(t,r)$ that can be extended to a joint distribution of $(t,r,y)$ satisfying
Assumptions \ref{assumption:reference_performance},
\ref{assumption:prevalence}, and \ref{assumption:everything_goes}. By
\Cref{prop:primitive-dilation-null}, the null set is
\begin{equation*}
\mathcal P_{0,s}
:=
\left\{
\Vec p\in\mathcal P_s:
g_L(\Vec p;s)\ge0 \text{ and }
g_U(\Vec p;s)\ge0
\right\}.
\end{equation*}

Thus, the hypotheses can equivalently be written as
$H_0:\Vec p\in\mathcal P_{0,s}$ and
$H_1:\Vec p\in\mathcal P_s\setminus\mathcal P_{0,s}$. To operationalize the test, 
recall that the observed data 
$W_i=(t_i,r_i)\in\{0,1\}^2$ are independent and identically distributed
for $i=1,\ldots,n$. Define the counts and count margins:
\[
N_{jk}:=\sum_{i=1}^n\mathbbm{1}\{t_i=j,r_i=k\},
\qquad
N_{j\cdot}:=N_{j0}+N_{j1},
\qquad
N_{\cdot k}:=N_{0k}+N_{1k}.
\]
The sample analogues of the identified prevalence and the two gaps are
\begin{align*}
\widehat P_s(y=1)
&:=
\frac{N_{\cdot1}/n+s_0-1}{s_0+s_1-1},
\\
\widehat g_L
&:=
\frac{N_{11}}n
-\widehat P_s(y=1)
\left(\frac{N_{1\cdot}}n+s_1-1\right), \text{ and}
\\
\widehat g_U
&:=
\frac{N_{01}}n
-\widehat P_s(y=1)
\left(s_1-\frac{N_{1\cdot}}n\right).
\end{align*}
Define the test statistic
\begin{equation}
H_n(N)
:=
n^2(s_0+s_1-1)
\max\bigl\{0,-\widehat g_L,-\widehat g_U\bigr\}.
\label{eq:exact-band-statistic}
\end{equation}

Large values indicate a violation of at least one gap inequality. The positive factor $n^2(s_0+s_1-1)$ cancels the frequency denominators in the sample gaps which is used to simplify computation. See Appendix \ref{app:computation} for details. The performance study sample is $
N:=(N_{00},N_{01},N_{10},N_{11})
\sim\operatorname{Multinomial}
\left(n;\Vec{p}\right)$, by Assumption \ref{random_sampling_assumption}. By Assumption \ref{assumption:everything_goes}, $p_{\cdot k}>0$ for $k\in\{0,1\}$, and the multinomial law has the exact factorization
\begin{align}\label{eq:factorization}
\begin{split}
   N_{\cdot1}
&\sim
\operatorname{Bin}(n,p_{\cdot1}),
\\
N_{11}\mid N_{\cdot1}
&\sim
\operatorname{Bin}
\left(
N_{\cdot1},
\frac{p_{11}}{p_{\cdot1}}
\right),
\\
N_{10}\mid N_{\cdot1}
&\sim
\operatorname{Bin}
\left(
n-N_{\cdot1},
\frac{p_{10}}{p_{\cdot0}}
\right), 
\end{split}
\end{align}
where $N_{11}\independent N_{10}\mid N_{\cdot1}$. For a given $\Vec{p}\in\mathcal{P}_{s}$, the probability of observing a hypothetical sample $N'\sim N$
whose statistic is at least as large as the observed statistic $H_n(N)$ is thus
\begin{align}
p_{\Vec{p}}(N):=& P_{\Vec{p}}\left(\left\{
H_n(N')\ge H_n(N)
\right\}\right)\nonumber\\
=&
\sum_{n'_{\cdot1}=0}^{n}
\binom{n}{n'_{\cdot1}}
p_{\cdot1}^{\,n'_{\cdot1}}
p_{\cdot0}^{\,n-n'_{\cdot1}}
\sum_{n'_{10}=0}^{n-n'_{\cdot1}}
\binom{n-n'_{\cdot1}}{n'_{10}}
\left(\frac{p_{10}}{p_{\cdot0}}\right)^{n'_{10}}
\left(\frac{p_{00}}{p_{\cdot0}}\right)^{
n-n'_{\cdot1}-n'_{10}}
\nonumber\\
&\times
\sum_{n'_{11}=0}^{n'_{\cdot1}}
\binom{n'_{\cdot1}}{n'_{11}}
\left(\frac{p_{11}}{p_{\cdot1}}\right)^{n'_{11}}
\left(\frac{p_{01}}{p_{\cdot1}}\right)^{
n'_{\cdot1}-n'_{11}}
\mathbbm{1}
\left\{
H_n(n')\ge H_n(N)
\right\},
\label{eq:fixed-distribution-binomial-tail}
\end{align}
where $n'_{\cdot k } = n'_{0k}+n'_{1k}$, $n'_{j \cdot } = n'_{j0}+n'_{j1}$ and $n'
=
\left(
n'_{00},n'_{01}, n'_{10}, n'_{11}
\right).$ 

When $\Vec p\in \mathcal{P}_{0,s}$ is the true distribution, this is a finite-sample valid $p$-value. However, $\Vec p$ is unknown and represents a nuisance parameter. Maximizing \eqref{eq:fixed-distribution-binomial-tail} over
$\Vec p\in\mathcal P_{0,s}$ would yield a valid least-favorable $p$-value,
but could be overly conservative. Following \citet{berger1994p}, we instead
localize the supremum to those null distributions that lie in a confidence
set for $\Vec p$. Specifically, let $C_N$ be a finite-sample
$1-\beta_n$ confidence set satisfying
\begin{equation}\label{eq:conf_int}
\inf_{\Vec{p}\in\mathcal P_s}
P_{\Vec{p}}\!\left(\Vec{p}\in C_N\right)
\geq 1-\beta_n
\end{equation}
for $0<\beta_n<\alpha<1$. The corresponding $p$-value is then
\begin{equation}
p_{\mathrm{BB}}(N)
:=\min
\left\{
1,\,
\beta_n+
\sup_{\Vec{p}\in C_N\cap \mathcal{P}_{0,s}}
p_{\Vec{p}}(N)
\right\},
\label{eq:berger-boos-pvalue}
\end{equation}
with the convention $
\sup_{\Vec p\in\emptyset}p_{\Vec p}(N):=0. $
Thus, if $C_N\cap\mathcal P_{0,s}=\emptyset$, then
$p_{\mathrm{BB}}(N)=\beta_n$. The test rejects $H_0$ if and only if $p_{\mathrm{BB}}(N)\leq \alpha$. Such a test is finite-sample valid, as shown by the following result.

\begin{theorem}
\label{th:finite-sample-validity}
Maintain Assumptions \ref{random_sampling_assumption},
\ref{assumption:reference_performance}, \ref{assumption:prevalence}, and
\ref{assumption:everything_goes}, and suppose $s\ne(1,1)$. Then, for every finite $n$ and every \(\gamma \in [0,1]\),
\[
\sup_{\Vec p\in\mathcal P_{0,s}}
P_{\Vec{p}}
\left\{
p_{\mathrm{BB}}(N)\le\gamma
\right\}
\leq
\gamma.
\]
Consequently, the test that rejects when
$p_{\mathrm{BB}}(N)\le\alpha$ has size at most $\alpha$.
\end{theorem}

The result above holds for any confidence set $C_N$ satisfying \eqref{eq:conf_int}. For example, one may construct $C_N$ using the level-set approach with minimum average volume \citep{chafai2009confidence, malloy2020optimal}. Given the factorization in \eqref{eq:factorization}, we instead use a computationally convenient construction based on \citet{clopper1934use}.

For $0\le k\le m$ and $0<\gamma<1$, let
$\operatorname{CP}_{1-\gamma}(k;m)$ denote the $1-\gamma$ equal-tailed
Clopper--Pearson confidence interval for a binomial probability based on $k$
successes in $m$ trials. When $m=0$, define
$\operatorname{CP}_{1-\gamma}(0;0):=[0,1]$. When
$X\sim\operatorname{Bin}(m,p_x)$, for every $p_x\in[0,1]$, the Clopper--Pearson interval satisfies
\(
P_{p_x}
\left\{
p_x\in
\operatorname{CP}_{1-\gamma}(X;m)
\right\}
\ge 1-\gamma.
\)
Therefore, a valid candidate for $C_N$, using the Bonferroni correction $\beta_n/3$, is

\begin{equation}
C_N
:=
\left\{
\Vec{p}\in\mathcal P_s:\;
\begin{aligned}
p_{\cdot1}
&\in
\operatorname{CP}_{1-\beta_n/3}
\left(N_{\cdot1};n\right),\\
\frac{p_{11}}{p_{\cdot1}}
&\in
\operatorname{CP}_{1-\beta_n/3}
\left(N_{11};N_{\cdot1}\right), \text{ and}
\\
\frac{p_{10}}{p_{\cdot0}}
&\in
\operatorname{CP}_{1-\beta_n/3}
\left(N_{10};N_{\cdot0}\right)
\end{aligned}
\right\}.
\label{eq:confidence-localization}
\end{equation}

The first interval has noncoverage probability at most $\beta_n/3$. By the
conditional binomial laws in
\eqref{eq:factorization},
each of the other two intervals has conditional noncoverage probability at
most $\beta_n/3$ for every possible value of $N_{\cdot1}$. Averaging over
$N_{\cdot1}$ gives the same unconditional bound. The union bound therefore
gives
\begin{equation*}
\inf_{\Vec p\in\mathcal P_s}
P_{\Vec p}\{\Vec p\in C_N\}
\ge1-\beta_n.
\end{equation*}

We relegate the details of implementation of the test to Appendix
\ref{app:computation}. There, we show that, for each fixed hypothetical pair
$(n'_{\cdot1},n'_{10})$, the event $H_n(n')\geq H_n(N)$ is the union of a
lower and an upper binomial tail in $n'_{11}$. This eliminates the innermost
sum in \eqref{eq:fixed-distribution-binomial-tail}, reducing the calculation
from $\binom{n+3}{3}$ individual tables to $(n+1)(n+2)/2$ pairs. We then
describe a branch-and-bound algorithm for maximizing this probability over $C_N\cap\mathcal P_{0,s}$. The algorithm establishes lower and
upper bounds on the supremum. It rejects when the upper bound implies that
the Berger--Boos $p$-value is no greater than $\alpha$, and it does not reject
when the lower bound implies that the $p$-value exceeds $\alpha$. If neither
conclusion follows, it subdivides the remaining regions to tighten the
bounds, and proceeds recursively.

\subsection{Simulations} \label{subsect:simulations}

In this section, we compare our test with two alternatives based on general-purpose methods for moment inequalities: a projection test using \citet{romano2014practical} and the minimum resampling test of \citet{bugni2017inference} for subvector inference. Both rely on the characterization of the null in \Cref{th:simpler_inference}, $H_0: \exists \theta\in\Theta \text{ s.t. } \theta_1+\theta_0=1$. The projection test checks whether the confidence set for $\theta$ constructed using \citet{romano2014practical} intersects the antidiagonal $\{\theta\in[0,1]^2:\theta_1+\theta_0=1\}$. The second applies the subvector minimum resampling test of \citet{bugni2017inference} directly to this $H_0$.\footnote{\citet[Theorem 1]{obradovic-2022} shows that \citet{romano2014practical} yields a confidence set for $\theta$ that is uniformly valid in level over a large class of DGPs. An earlier version of this paper showed that the minimum resampling test is also uniformly valid under mild conditions.}

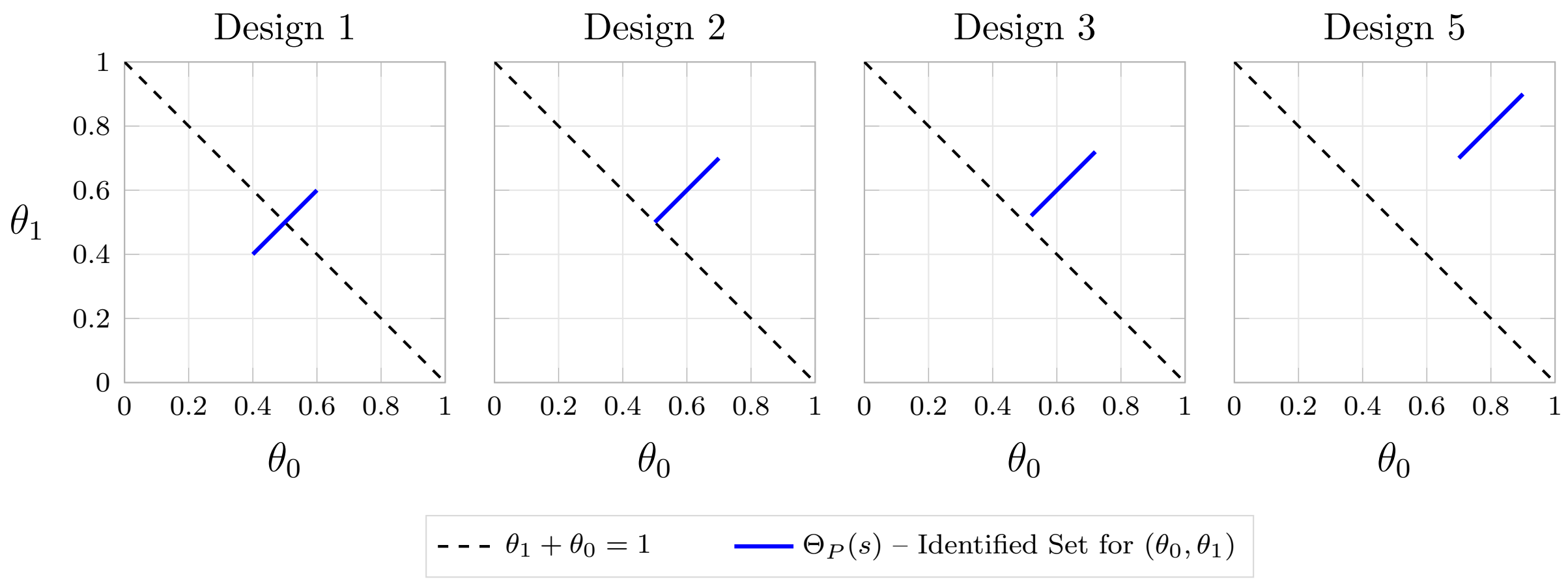
\begin{figure}[htbp]
\centering

\begin{tikzpicture}

\begin{groupplot}[
    group style={
        group size=4 by 1,
        horizontal sep=0.5cm
    },
    width=0.29\textwidth,
    height=0.29\textwidth,
    xmin=0, xmax=1,
    ymin=0, ymax=1,
    xtick={0,0.2,0.4,0.6,0.8,1},
    ytick={0,0.2,0.4,0.6,0.8,1},
    tick label style={font=\scriptsize},
    grid=major,
    grid style={gray!20},
    axis line style={gray!60},
    tick style={gray!60},
    xlabel={$\theta_0$},
    ylabel={$\theta_1$},
    ylabel style={rotate=-90},
    title style={font=\small,yshift=-2mm}
]

% ---------- Design 1 ----------
\nextgroupplot[
    title={Design 1},
    legend to name=designlegend,
    legend columns=2,
    legend style={
        draw=gray!30,
        fill=white,
        font=\scriptsize,
        inner sep=3pt,
        /tikz/every even column/.append style={column sep=0.75cm}
    }
]

\addplot[
    black,
    thick,
    dashed,
    domain=0:1,
    samples=2
] {1-x};
\addlegendentry{$\theta_1+\theta_0=1$}

\addplot[
    blue,
    very thick
]
coordinates {
    (0.40,0.40)
    (0.60,0.60)
};
\addlegendentry{$\Theta_P(s)$ -- Identified Set for $(\theta_0,\theta_1)$}

% ---------- Design 2 ----------
\nextgroupplot[
    title={Design 2},
    title style={font=\small},
    ylabel={},
    yticklabels=\empty
]

\addplot[
    black,
    thick,
    dashed,
    domain=0:1,
    samples=2
] {1-x};

\addplot[
    blue,
    very thick
]
coordinates {
    (0.50,0.50)
    (0.70,0.70)
};

% ---------- Design 3 ----------
\nextgroupplot[
    title={Design 3},
    title style={font=\small},
    ylabel={},
    yticklabels=\empty
]

\addplot[
    black,
    thick,
    dashed,
    domain=0:1,
    samples=2
] {1-x};

\addplot[
    blue,
    very thick
]
coordinates {
    (0.52,0.52)
    (0.72,0.72)
};

% ---------- Design 5 ----------
\nextgroupplot[
    title={Design 5},
    title style={font=\small},
    ylabel={},
    yticklabels=\empty
]

\addplot[
    black,
    thick,
    dashed,
    domain=0:1,
    samples=2
] {1-x};

\addplot[
    blue,
    very thick
]
coordinates {
    (0.70,0.70)
    (0.90,0.90)
};

\end{groupplot}

% ---------- Common legend ----------
\node[anchor=north]
    at ([yshift=-12mm]
        $(group c1r1.south)!0.5!(group c4r1.south)$)
    {\pgfplotslegendfromname{designlegend}};

\end{tikzpicture}

\caption{Identified sets for designs in \Cref{tab:simulation_designs}.
The identified set in Design 4 lies between those in Designs 3 and 5,
and is omitted.}
\label{fig:simulation_designs}

\end{figure}

We consider five DGPs defined in \Cref{tab:simulation_designs} and fix the reference performance at $s=(0.9,0.9)$. The corresponding identified sets are shown in \Cref{fig:simulation_designs}. The first two designs yield a dilation-inducing test and are used to assess validity. In the first, the antidiagonal intersects $\Theta_P(s)$ at the midpoint of the identified set. All tests are expected to underreject in this case, since the sample identified set is likely to intersect the antidiagonal, leading to non-rejection. In the second design, the antidiagonal intersects $\Theta_P(s)$ at its boundary, illustrating how conservative the tests are. The remaining DGPs do not yield a dilation-inducing test and are used to evaluate power. The third is a perturbation of the second, with $\Theta_P(s)$ lying just above the antidiagonal. The fourth and fifth move $\Theta_P(s)$ progressively farther away from the antidiagonal.

\begin{table}[ht!]
    \centering
    \caption{Population designs.}
    \label{tab:simulation_designs}
    \begin{tabular}{@{}cccccc@{}}
        \toprule
        Design & $p_{00}$ & $p_{01}$ & $p_{10}$ & $p_{11}$ & $H_0$ \\
        \midrule
        1 & $25\%$ & $25\%$ & $25\%$ & $25\%$ & True  \\
        2 & $30\%$ & $20\%$ & $20\%$ & $30\%$ & True \\
        3 & $31\%$ & $19\%$ & $19\%$ & $31\%$ & False \\
        4 & $35\%$ & $15\%$ & $15\%$ & $35\%$ & False \\
        5 & $40\%$ & $10\%$ & $10\%$ & $40\%$ & False \\
        \bottomrule
    \end{tabular}
\end{table}

We use three different sample sizes: $n \in \{50, 100, 500\}$. These sample
sizes are relatively small but are typical for the applications in mind. Each
reported entry is based on $1{,}000$ Monte Carlo simulations, and the
significance level is $5\%$. The implementation of our test uses
$\beta_n=0.001$ for all $n$, following \citet{berger1994p}.

\begin{table}[ht!]
	\centering
	\setlength{\tabcolsep}{4pt}
	\caption{Simulation results: Observed rejection probabilities.}
	\label{tab:simulation}
	\begin{tabular}{@{}ccccccc@{}}
		\toprule
		\multicolumn{2}{c}{} & Design 1 & Design 2 & Design 3 & Design 4 & Design 5 \\ \midrule
		\multicolumn{2}{c}{$H_0$} & True & True & False & False & False \\ \midrule
		\multirow{3}{*}{$n=50$}
		& RSW & $0\%$ & $0\%$ & $0\%$ & $0.1\%$ & $4.3\%$ \\
		& BCS & $0\%$ & $0.8\%$ & $3.9\%$ & $22\%$ & $76\%$ \\
		& BB  & $0.2\%$ & $2.4\%$ & $6.1\%$ & $35.4\%$ & $90.7\%$ \\ \cline{3-7}
		\multirow{3}{*}{$n=100$}
		& RSW & $0\%$ & $0\%$ & $0\%$ & $0.1\%$ & $19\%$ \\
		& BCS & $0\%$ & $1\%$ & $3.1\%$ & $43\%$ & $100\%$ \\
		& BB  & $0\%$ & $3.3\%$ & $7.3\%$ & $61.7\%$ & $99.2\%$ \\ \cline{3-7}
		\multirow{3}{*}{$n=500$}
		& RSW & $0\%$ & $0\%$ & $0\%$ & $28\%$ & $100\%$ \\
		& BCS & $0\%$ & $1.3\%$ & $9.2\%$ & $99\%$ & $100\%$ \\
		& BB  & $0\%$ & $1.9\%$ & $13.9\%$ & $99.2\%$ & $100\%$ \\ \bottomrule
		\multicolumn{7}{p{11.8cm}}{\footnotesize $1{,}000$ Monte Carlo iterations per design, sample size, and method. RSW and BCS denote the tests based on \cite{romano2014practical} and \cite{bugni2017inference}, respectively. BB denotes our test.}
	\end{tabular}
\end{table}

\Cref{tab:simulation} reports the results. No test overrejects in either of the first two designs. Our test (BB) is conservative, as expected given the discreteness of the data, but is strictly less conservative in Design 2 and weakly less conservative in Design 1 across all $n$. In terms of power, BB dominates nearly uniformly. Compared with BCS, which itself dominates RSW, BB substantially raises rejection rates at $n=50$ and $n=100$ in Designs 3 and 4. At $n=500$, it substantially raises power in Design 3 and slightly raises it in Design 4. In Design 5, BB substantially raises power at $n=50$, slightly underperforms BCS at $n=100$, and performs equally well as the other tests at $n=500$.

\section{Empirical Application: CT chest scans for COVID-19 detection} \label{sect:applications}

Early in the COVID-19 pandemic, highly influential work by \citet{ai2020correlation} endorsed the use of CT chest scans for COVID-19 detection. The authors concluded that chest CT scans could serve as a primary detection tool, contributing to their adoption in clinical practice at some hospitals \citep{de2020radiological}. We revisit their data and cannot reject the hypothesis that the test was dilation-inducing.

The performance study data, shown in \autoref{tab:aietal_data}, were collected at a hospital in Wuhan, China, in early 2020. Here, the index test $t$ is the radiologist's interpretation of a chest CT scan, and the underlying health status $y$ indicates whether the patient has COVID-19. The test was evaluated against an RT-PCR test, which served as the reference $r$. The recommendation for use as a primary detection tool was based on high estimated apparent sensitivity, $\tilde{\theta}_1=P(t=1\mid r=1)$, i.e.\ sensitivity relative to the reference test. The estimate is $580/601=96.5\%$, also exceeding the $80\%$ apparent sensitivity threshold previously used by the FDA for emergency use authorization of COVID-19 tests evaluated against the same type of reference test. However, as shown in \Cref{subsect:example_dil}, a test may surpass such thresholds and still be dilation-inducing, even when the reference test is highly accurate.

\begin{table}[ht!]
	\centering
	\caption{Data from \cite{ai2020correlation}.}
	\label{tab:aietal_data}
	\begin{tabular}{@{}cccc@{}}
		\toprule
		 & $r=0$ & $r=1$ & 	\\ \midrule
		$t=0$ & $105$ & $21$ & $126$\\
		$t=1$ & $308$ & $580$ & $888$\\ \midrule
		 & $413$ & $601$ & $n=1014$ \\
		\bottomrule
	\end{tabular}
\end{table}

RT-PCR tests are typically considered to be highly accurate, but are known not to be perfectly sensitive (see, e.g., \citealp{arevalo2020false}). Following \citet{kucirka2020variation} and \cite{kanji2021false}, we assume $s=(1, 0.9)$. For additional details regarding the choice of $s$, see \citet[Section 5.1]{obradovic-2022}.  \Cref{fig:example2} depicts the plug-in estimate of the identified set $\Theta_P(s)$. Appendix \ref{app:extensions} discusses the robustness of findings to different assumed values of $s$. The estimated set intersects the antidiagonal, yielding the statistic $H_n(N) = 0$ and the corresponding $p$-value $p_{BB}(N) = 1$. Hence, the finite-sample valid test cannot reject that $t$ is dilation-inducing at any conventional level of significance. 

Under the maintained assumptions, this evidence is consistent with CT scans being dilation-inducing, despite being recommended for use based on conventional performance measures, as well as passing previously used thresholds set by the FDA for test approval. \autoref{fig:aietal_pv} illustrates the resulting dilation when the estimated set is used to update an example pre-test probability $\pi=\nicefrac{1}{3}$. A positive CT scan yields relatively little ambiguity, as reflected by the size of the post-test probability set. A negative CT scan yields substantially more. Nevertheless, the two sets always include $\pi$ and overlap, so the clinician faces dilation. The same occurs for any $\pi\in(0,1)$.

\begin{figure}[ht!]
	\centering

		\begin{tikzpicture}[line cap=round, line join=round, x=5cm, y=5.0cm, scale=0.75, dot/.style={circle,fill=black,minimum size=4pt,inner sep=0pt,  outer sep=-1pt}]
			\draw [->,color=black] (0,-0.025) -- (0,1.05) node [above] {\footnotesize $Q(y=1)$};
			\draw [->,color=black] (0.67,-0.025) -- (0.67,1.05) node [above] {\footnotesize $Q(y=1|t)$};
			\draw [-,color=black] (0.015,1) -- (-0.015,1) node [left] {$1$};
			\draw [-,color=black] (0.6815,1) -- (.655,1);
			\draw [-,color=black] (0.015,0) -- (-0.015,0) node [left] {$0$};
			\draw [-,color=black] (0.6815,0) -- (.655,0);
			\draw[black, fill=black, opacity=1] (-0.03,0.335) rectangle (0.03,0.331);
			\draw[preaction={fill, blue, opacity=0.35}, rounded corners=2pt, pattern=horizontal lines,pattern color=black] (0.64,0.3280) rectangle (0.7,0.4101);
			\draw[preaction={fill, red, opacity=0.35}, rounded corners=2pt, pattern=north west lines,pattern color=black] (0.64,0.0493) rectangle (0.7,0.3732);
			\draw [dotted,color=black!25] (-0.05,0.3333) -- (.7,0.3333);
			\draw[dashed,color=black!33] (0.0,1) -- (0.67,1);
			\draw[dashed,color=black!33] (0.0,0) -- (0.67,0);
			\draw [->,>=stealth,out=15,in=180,color=blue, thick] (0.1,0.3333) to node [yshift=5, midway, above, blue] {$t=1$} (0.6,0.38);
			\draw [->,>=stealth,out=-45,in=180,color=red, thick] (0.1,0.3333) to node [yshift=-5, midway, below, red] {$t=0$} (0.6,0.15);
			\node[left] at (-0.03,0.3333) {$\pi$};
			\node[right] at (0.7,0.41) {$V_1$};
			\node[right] at (0.7,0.15) {$V_0$};
		\end{tikzpicture}

	\caption{Sets of post-test probabilities for the empirical distribution of \cite{ai2020correlation} assuming $\pi=\nicefrac{1}{3}$.}
	\label{fig:aietal_pv}
\end{figure}

Being dilation-inducing means that updating on the scan results only introduces ambiguity about the patient's true health status, but it need not reflect an inherent failing of the imaging modality. Our findings concern the test as evaluated in \citet{ai2020correlation} under the maintained assumptions. Whether the test is dilation-inducing depends on both the joint distribution of index and reference test results in the study and the information available about the reference test's true performance. Consequently, our findings need not extend to other studies or clinical settings.

The application illustrates how the same performance-study data used to support a recommendation for clinical use can also be used to assess the potential for dilation before implementation. Such an assessment could inform decisions about test approval and use, particularly in light of subsequent work reporting limited diagnostic performance and recommending against routine CT screening, which had been discontinued in certain settings \citep{lee2022diagnostic}. The method could also be applied to the performance evidence supporting the more selective diagnostic uses of CT imaging retained in WHO guidance \citep{who2022imaging}, potentially reducing unnecessary testing and associated harm to patients.

\section{Conclusion and Policy Implications} \label{sect:conclusion}

This paper highlights that standard clinical practices can be conducive to dilation, a property typically viewed as undesirable in models of ambiguity. Moreover, dilation may arise from a ubiquitous feature of test performance studies, even for tests that appear to perform well by conventional standards. We thus conclude with a direct recommendation to regulators and researchers: whenever an imperfect reference test is used, one should evaluate new diagnostic tests based on their true performance rather than apparent performance alone and, in doing so, verify whether they are dilation-inducing. This may reveal that a seemingly well-performing test is in fact \textit{only} introducing ambiguity about the patient's true health status, and not approving it may thereby reduce the losses stemming from implementing such tests. To facilitate verification, we provide an equivalent characterization of dilation-inducing tests, a finite-sample valid hypothesis test, and tools for sensitivity analysis when reference-test performance is not known exactly.

\bibliographystyle{ecta} 
\bibliography{informative_tests}

\appendix

%%%%%%%%%%%%%%%%%%%%%%%%%%%%%%%%%%%%%%%%%%%%%%%%%%%%%%%%%%%%%%%%%%
%%%%%%%%%%%%%%%%%%%%%%%%%%%%%%%%%%%%%%%%%%%%%%%%%%%%%%%%%%%%%%%%%%
%%%%%%%%%%%%%%%%%%%%%%%%%%%%%%%%%%%%%%%%%%%%%%%%%%%%%%%%%%%%%%%%%%
%%%%%%%%%%%%%%%%%%%%%%%%%%%%%%%%%%%%%%%%%%%%%%%%%%%%%%%%%%%%%%%%%%
%%%%%%%%%%%%%%%%%%%%%%%%%%%%%%%%%%%%%%%%%%%%%%%%%%%%%%%%%%%%%%%%%%
%%%%%%%%%%%%%%%%%%%%%%%%%%%%%%%%%%%%%%%%%%%%%%%%%%%%%%%%%%%%%%%%%%
%%%%%%%%%%%%%%%%%%%%%%%%%%%%%%%%%%%%%%%%%%%%%%%%%%%%%%%%%%%%%%%%%%
\appendix

\counterwithin{lemma}{section}
\renewcommand{\thelemma}{A\arabic{lemma}}
%%%%%%%%%%%%%%%%%%%%%%%%%%%%%%%%%%%%%%%%%%%%%%%%%%%%%%%%%%%%%%%%%%
%%%%%%%%%%%%%%%%%%%%%%%%%%%%%%%%%%%%%%%%%%%%%%%%%%%%%%%%%%%%%%%%%%

\section{Nomograms}\label{sect:nomogram}

In practice, clinicians frequently use physical or online versions of Fagan's nomogram, originally introduced by \citet{fagan1975nomogram}, to facilitate bedside calculation of post-test probabilities. The original nomogram takes as inputs $\pi$ and the positive and negative likelihood ratios, $LR^+$ and $LR^-$, obtained from $\theta$. These are defined as
\begin{equation*}
    LR^+ := \frac{\theta_1}{1-\theta_0} 
    \quad \text{ and } \quad
    LR^- := \frac{1-\theta_1}{\theta_0},
\end{equation*}
respectively.

\begin{figure}[ht!]
    \begin{subfigure}[b]{0.45\textwidth}
        \begin{center}
        \includegraphics[width=\textwidth]{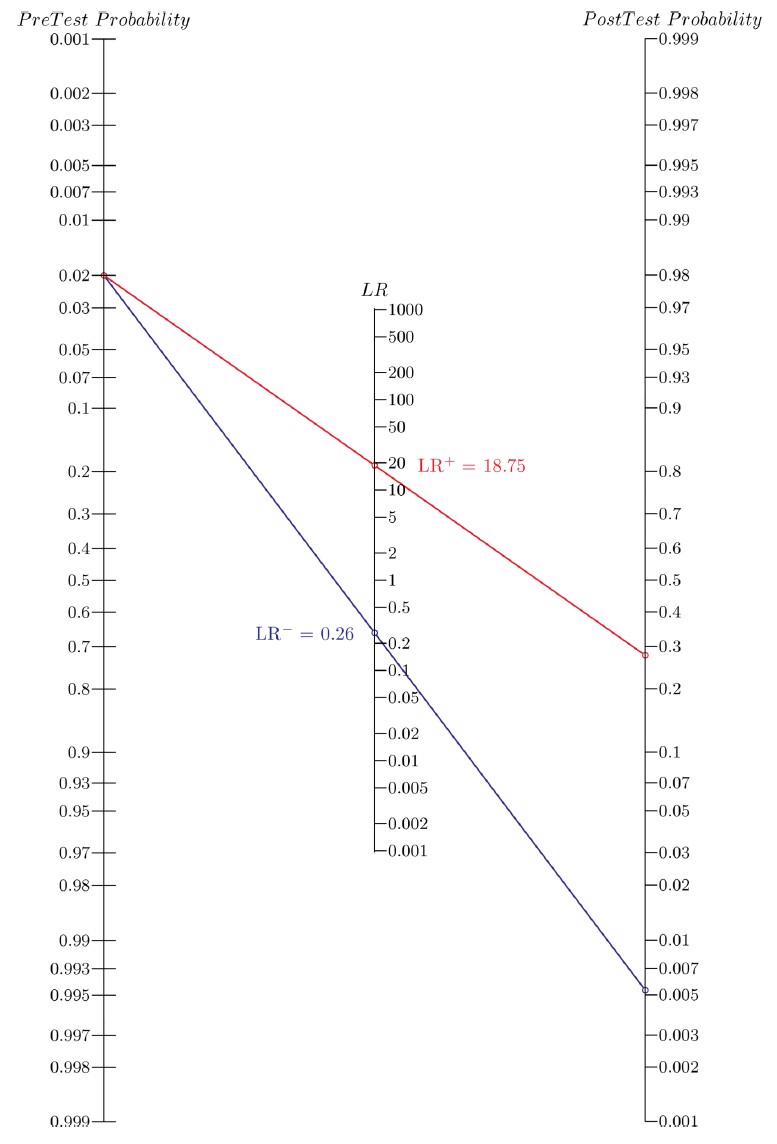}
        \caption{Fagan's nomogram.}
        \label{fig:fagan1}
        \end{center}
    \end{subfigure}
    \hfill
    \begin{subfigure}[b]{0.45\textwidth}
        \begin{center}
        \includegraphics[width=\textwidth]{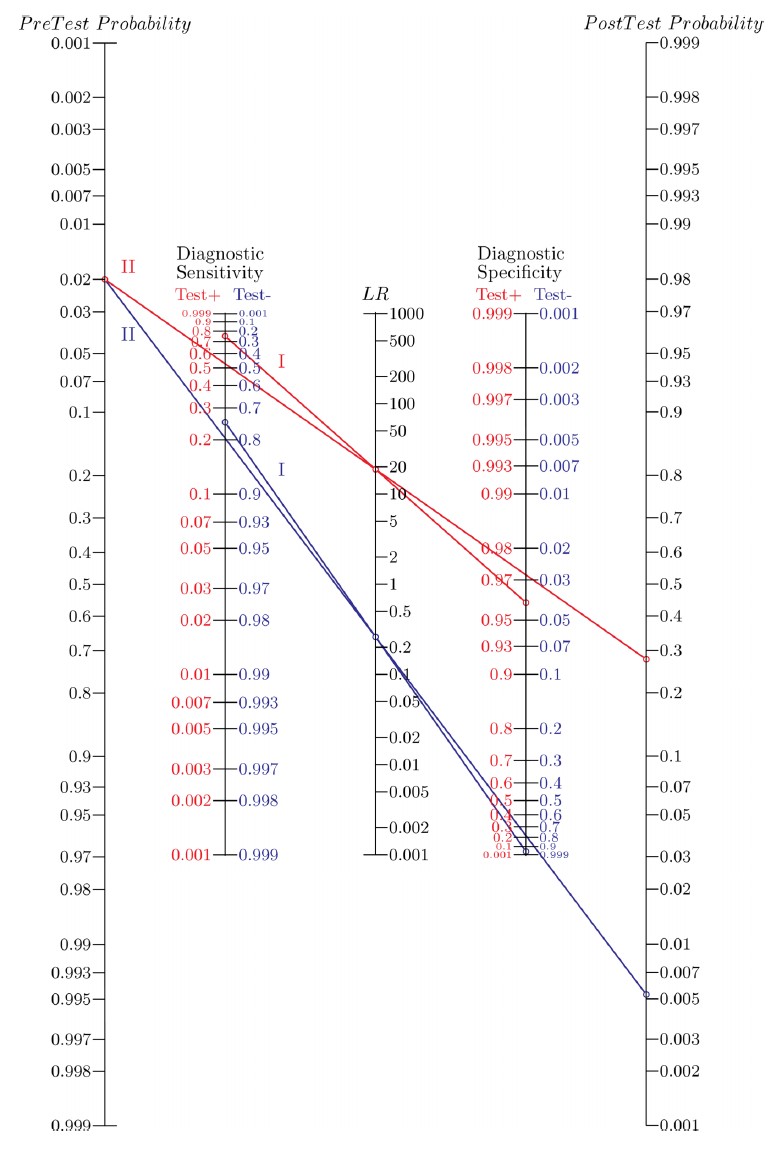}
        \caption{Two-step Fagan's nomogram.}
        \label{fig:fagan2}
        \end{center}
    \end{subfigure}
    \caption{Fagan's and two-step Fagan's nomogram calculation example from \citet{caraguel2013two}.}
   \caption*{\justifying{An MRI screening test for breast cancer with $\theta=(0.96,0.75)$ and $\pi=0.02$ is shown. In \autoref{fig:fagan1}, $LR^+$ and $LR^-$ are calculated and used as inputs. The red and blue lines then yield the positive and negative post-test probabilities $PPP=0.28$ and $NPP=0.006$, respectively. In \autoref{fig:fagan2}, the likelihood ratios are first obtained graphically from $\theta$: $LR^+=18.75$ (red line, I) and $LR^-=0.26$ (blue line, I). These are then used to obtain $PPP=0.28$ (red line, II) and $NPP=0.006$ (blue line, II).}}
    \label{fig:fagan-both}
\end{figure}

The clinician draws a straight line connecting the assessed $\pi$ to the appropriate likelihood ratio to obtain the corresponding post-test probability, as shown in \autoref{fig:fagan1}. Because diagnostic test performance studies more commonly focus on $\theta$ than likelihood ratios, two-step Fagan's nomograms were developed to avoid requiring clinicians to compute the ratios themselves. The clinician first draws a line connecting the relevant $\theta_0$ and $\theta_1$ values to obtain the corresponding likelihood ratio. A second line then maps that ratio into the post-test probability, as shown in \autoref{fig:fagan2}.

\subsection{Nomograms and Ambiguous Parameters}\label{subsect:nomogram_ambig}

\begin{figure}[ht!]
\begin{center}
    \includegraphics[width=0.75\textwidth]{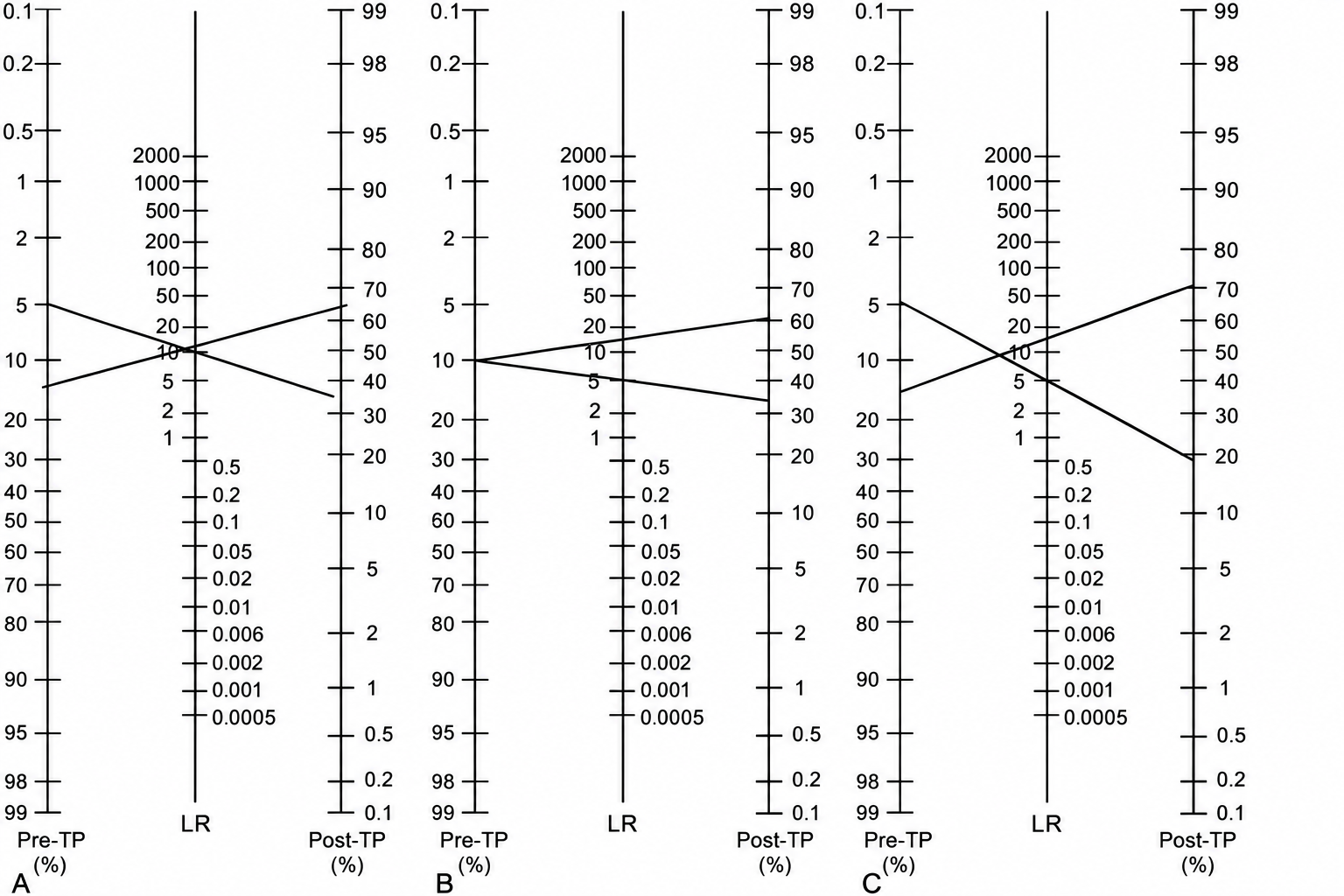}
\caption{Fagan's nomograms with ambiguous $\pi$ and/or $\theta$. Examples from \citet{srinivasan2012propagation}.}
	\label{fig:fagan_uncertainty}
\end{center}
     \caption*{\justifying{The Post-TP axis shows the bounds on the post-test probabilities obtained via the nomogram. Panels A, B and C depict ambiguity in $\pi$ only, $\theta$ only, and both $\pi$ and $\theta$, respectively.}}
\end{figure}

Clinicians typically incorporate ambiguity about $\pi$ and/or $\theta$ by constructing bounds on the post-test probability. These can be obtained directly from Bayes' theorem, as in \citet{manski-2021}, or equivalently through graphical tools. \citet{srinivasan2012propagation} provide examples, reproduced in \autoref{fig:fagan_uncertainty}. Panel A considers ambiguity in $\pi$ only. Bounds on the post-test probability are obtained by drawing lines through the appropriate likelihood ratio from the lowest and highest pre-test probabilities. Panel B considers ambiguity in $\theta$, and hence in the likelihood ratio, only. The bounds are obtained by drawing lines from the assessed $\pi$ through the endpoints of the likelihood-ratio bounds. Panel C considers ambiguity in both and proceeds analogously.

\section{Ambiguous Pre-Test Probability}\label{sect:pre_test_ambiguity}

In the main analysis, we assume that the clinician assesses a unique pre-test probability $\pi$. We now extend the characterization to settings in which the clinician cannot make a point assessment and instead knows only that the pre-test probability lies in a closed set $\Pi\subseteq(0,1)$. With a set of pre-test probabilities and a generic identified set $\Theta$ for $\theta$, we extend the post-test probability sets in \eqref{eq:post_test_definition} to
\begin{align}\label{eq:post_test_definition_set}
	V_j(\Theta;\Pi)  &:= \Big\{v_j(\theta;\pi) : \theta\in \Theta, \pi \in \Pi \Big\}\text{ for $j=0,1$}.
\end{align}
The notion of dilation extends naturally to this setting. Under the following generalization, the conclusions of \autoref{prop:dilation_characterization}, and hence the results based on it, continue to hold.
\begin{definition}\label{def:dilation_set}
	$t$ is dilation-inducing if for any closed nonempty $\Pi \subseteq (0,1)$, $\Pi \subsetneq V_1(\Theta;\Pi)$ and $\Pi \subsetneq V_0(\Theta;\Pi).$
\end{definition}

\begin{proposition}
\label{prop:dilation_characterization_set}
Let a dilation-inducing test be given by \Cref{def:dilation_set}. The index test is dilation-inducing if and only if there exist $\theta, \theta' \in \Theta$ such that $\theta_0+\theta_1=1$ and $\theta_0'+\theta_1'\neq 1$.
\end{proposition}

\section{Uncertainty about the reference test's performance} \label{app:extensions}

This section extends the analysis to account for uncertainty regarding the reference test's performance. First, we consider the case where the reference test performance $s$ is only known to lie within a set $\mathcal{S}$. This may be used to represent ambiguity regarding $s$ or as a sensitivity analysis of the results to the value of $s$.

\begin{assumptionS}{assumption:reference_performance}{(Generalized Reference Performance)}\label{assumption:reference_performance_S}
Sensitivity and specificity of the reference test are contained in a known nonempty set $\mathcal{S}\subset[0,1]^2$ such that $s_1+s_0>1$ for all $s\in\mathcal{S}$.
\end{assumptionS}

Although knowledge of $\mathcal{S}$ remains a non-trivial assumption, it is clearly a relaxation of \autoref{assumption:reference_performance}. The fact that the common assumption $s=(1,1)$ has been disputed for a wide range of reference tests indicates that, for many tests used as $r$, one can at least specify a set $\mathcal{S}$ of more credible values of $s$. We maintain that $s_0+s_1>1$ for every $s\in\mathcal{S}$, extending the earlier restriction to all admissible values.

Instead of requiring that each assumed value is feasible, we explicitly allow some values in $\mathcal{S}$ to be refuted by the observable population distribution. Specifically, values producing $P_s(y=1)\notin(0,1)$ contradict Assumption \ref{assumption:prevalence}. Keeping dependence on $P$ implicit, let
\[
\Tilde{\mathcal{S}}
:=
\{s\in\mathcal{S}:P_s(y=1)\in(0,1)\}
\]
denote the feasible values. The sharp identified set for $\theta$ is then
\begin{equation*}
    \Theta_P(\mathcal{S})
    :=
    \bigcup_{s\in\Tilde{\mathcal{S}}}\Theta_P(s).
\end{equation*}

By \Cref{prop:dilation_characterization}, the index test is dilation-inducing if and only if there exist $\theta,\theta'\in\Theta_P(\mathcal{S})$ such that $\theta_0+\theta_1=1$ and $\theta_0'+\theta_1'\neq1$. For every feasible $s$, the line specified in \eqref{eq:H_s} intersects the antidiagonal at the same point $\theta^\star(P)=(1-p_{1\cdot},p_{1\cdot})$. The union $\Theta_P(\mathcal{S})$ can therefore intersect the antidiagonal only at this point. Thus, equivalently, $t$ is dilation-inducing if and only if $\theta^\star(P)\in\Theta_P(\mathcal{S})$ and $\Theta_P(\mathcal{S})$ is non-singleton. Under full support, $\Theta_P(s)$ is non-singleton for every feasible $s\neq(1,1)$. The existence of one such value therefore ensures that $\Theta_P(\mathcal{S})$ is non-singleton. The result in \autoref{th:simpler_inference} generalizes directly and we have the following.

\begin{proposition}
\label{prop:primitive-dilation-nullS}
Maintain Assumptions \ref{assumption:reference_performance_S},
\ref{assumption:prevalence}, and \ref{assumption:everything_goes}, and suppose
$\Tilde{\mathcal{S}}\setminus\{(1,1)\}\neq \emptyset$.
The set $\Theta_P(\mathcal{S})$ can intersect the antidiagonal only at
$\theta^\star(P)=(1-p_{1\cdot},p_{1\cdot})$.
Moreover,
\[
\theta^\star(P)\in\Theta_P(\mathcal{S})
\quad\Longleftrightarrow\quad
\exists s\in\Tilde{\mathcal{S}},
\quad
g_L(\Vec p;s)\ge0
\quad\text{and}\quad
g_U(\Vec p;s)\ge0.
\]
Consequently, $t$ is dilation-inducing if and only if there exists
$s\in\Tilde{\mathcal{S}}$ such that both gap inequalities hold.
\end{proposition}

A hypothesis test for whether $t$ is dilation-inducing can then be
constructed by treating $s$ as an additional nuisance parameter in the
testing procedure outlined in \Cref{sect:inference}. For each candidate
$s\in\mathcal S$ and $\Vec p\in C_N\cap\mathcal P_{0,s}$, define
\[
p_{\Vec p,s}(N)
:=
P_{\Vec p}
\left\{
H_n(N';s)\geq H_n(N;s)
\right\},
\text{ with }
N'\sim\operatorname{Multinomial}(n;\Vec p).
\]
Thus, the quantity being maximized is the probability, under the candidate
distribution $\Vec p$, that a hypothetical sample produces an
$s$-specific statistic at least as large as the observed statistic. The
corresponding Berger--Boos $p$-value is
\[
p_{\mathrm{BB}}^{\mathcal S}(N)
=
\min\left\{
1,\,
\beta_n+
\sup_{\substack{
s\in\mathcal S\\
\Vec p\in C_N\cap\mathcal P_{0,s}
}}
p_{\Vec p,s}(N)
\right\}.
\]

Second, we characterize the reference-test performances under which the
index test is dilation-inducing. Researchers may then assess whether the true
performance of $r$ plausibly lies in this set. This reasoning leads to the
definition of a \emph{dilator set}.

\begin{definition}
For a population distribution $P$ and a set of candidate reference-test
performances $\mathcal S$, define the \emph{dilator set} as
\[
\mathcal D_P(\mathcal S)
:=
\left\{
s\in\mathcal S:
\exists
\theta,\theta'\in\Theta_P(s)
\text{ such that }
\theta_1+\theta_0=1\text{ and } \theta_1'+\theta_0'\neq 1
\right\}.
\]
\end{definition}

The dilator set is a population object: it depends on the population distribution $P$, while $\mathcal S$ specifies the reference-test performances considered by the researcher. For any candidate
$s$ under which $P$ is incompatible with the maintained model, we adopt the
convention that $\Theta_P(s)=\emptyset$.

This approach also accommodates settings in which the researcher prefers not to impose explicit assumptions on $s$, beyond mild regularity conditions. For example, they may set $\mathcal S = \mathcal{S}_>= \{s\in[0,1]^2: s_1+s_0>1\}$ which encompasses the case where $r$ is only assumed to be better than a coin toss.

 Under the maintained assumptions, the dilator set has a directly observable characterization. Combining \Cref{th:simpler_inference} with \Cref{prop:primitive-dilation-null} gives 
 \begin{equation*} \mathcal D_P(\mathcal S) = \left\{ s\in\mathcal S:	\Vec p\in\mathcal P_s,\quad s\neq(1,1),\quad g_L(\Vec p;s)\geq0,\quad g_U(\Vec p;s)\geq0 \right\}. 
 \end{equation*}

\begin{figure}[ht!] 
  \centering
  \includegraphics[width=0.7\textwidth]{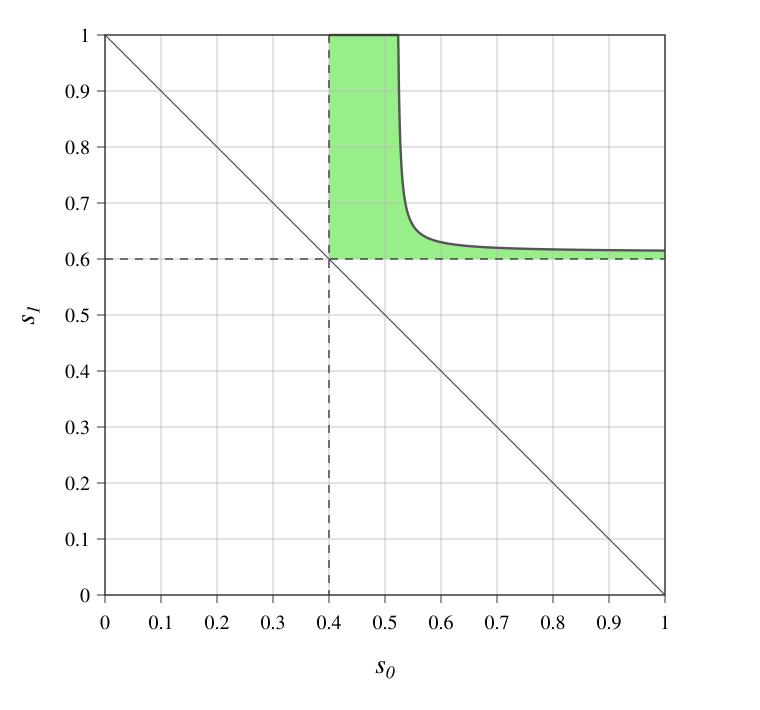}
  \caption{Dilator set $\mathcal{D}_P(\mathcal{S}_>)$ for the data-generating process given by \autoref{tab:policy_example}. The dashed lines correspond to $s_0=P(r=0)$ and $s_1=P(r=1)$.}
  \label{fig:dilator_policy}
\end{figure}

\noindent\autoref{fig:dilator_policy} shows an example of a dilator set using the DGP from \autoref{tab:policy_example}.

The population distribution $P$ is unknown. Let $\widehat P_N$ denote the empirical distribution defined by $\widehat P_N(t=j,r=k)=\frac{N_{jk}}{n}.$ 

The plug-in dilator set is 
\(
\widehat{\mathcal D}_N(\mathcal S) := \mathcal D_{\widehat P_N}(\mathcal S). \) It provides a direct sample analogue of the population dilator set. One can also use this approach to formalize a nonrejection region across $s$ with pointwise coverage, as a form of sensitivity analysis. This may be done by applying the test developed in \Cref{sect:inference} separately at each candidate $s\in\mathcal S$. Let $p_{\mathrm{BB}}(N;s)$ denote the Berger--Boos $p$-value for the null hypothesis $H_0(s): \Vec p\in\mathcal P_{0,s}$. The pointwise nonrejection region is 

\begin{equation} 
\widehat{\mathcal D}^{\,\mathrm{pt}}_{N,1-\alpha}(\mathcal S) := \left\{ s\in\mathcal S: p_{\mathrm{BB}}(N;s)>\alpha \right\}. 
\label{eq:pointwise-dilator-region} 
\end{equation} 

For every fixed $s\in\mathcal D_P(\mathcal S)$, finite-sample validity of the test implies 
\[ 
P_{\Vec p}\left\{ s\in \widehat{\mathcal D}^{\,\mathrm{pt}}_{N,1-\alpha}(\mathcal S) \right\} \geq1-\alpha. 
\] 

Values in the region are those for which dilation is not rejected. This exercise differs from treating $s$ as a nuisance parameter and reporting a single test over $\mathcal S$. The latter maximizes the tail probability jointly over all candidate $s\in\mathcal S$ and produces one worst-case $p$-value. By contrast, \eqref{eq:pointwise-dilator-region} retains the candidate-by-candidate conclusions that may be used as a sensitivity analysis to changes in assumed $s$.

\subsection{Robustness of findings in
  \texorpdfstring{\Cref{sect:applications}}{Section 6}}

\begin{figure}[ht!] 
  \centering
  \includegraphics[width=0.7\textwidth]{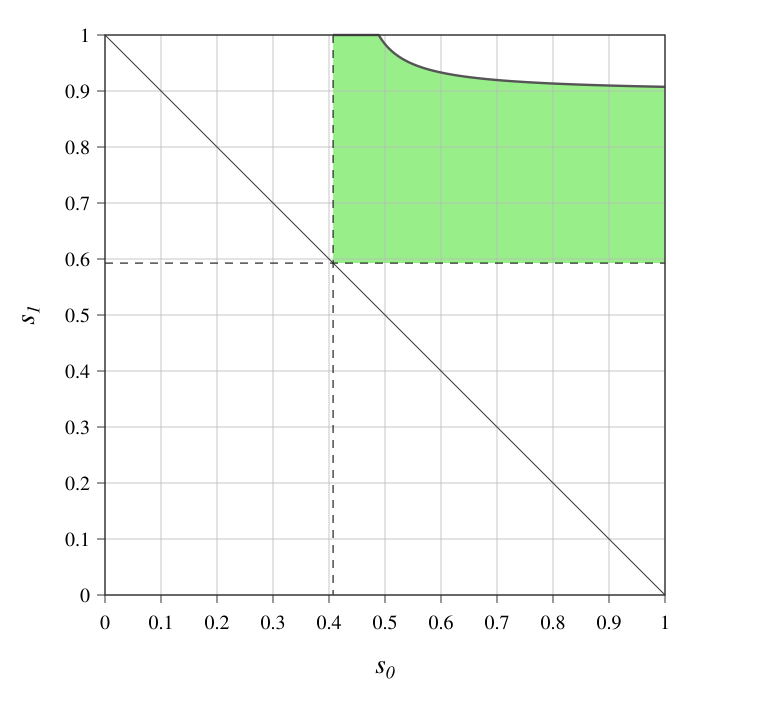}
  \caption{Plug-in dilator set $\widehat{\mathcal D}_{N}(\mathcal S_>)$ for the sample given by \autoref{tab:aietal_data}. The dashed lines correspond to $s_0=\widehat P_N(r=0)$
and $s_1=\widehat P_N(r=1)$.}
  \label{fig:dil_set_robustness}
\end{figure}

The findings in the empirical application hinge on the chosen value of $s$. We therefore conduct a sensitivity analysis using the tools discussed above. 
The plug-in dilator set $\widehat{\mathcal D}_{N}(\mathcal S_>)$, shown in \Cref{fig:dil_set_robustness}, indicates that values $s_1<0.9$ can also make $t$ dilation-inducing. The pointwise nonrejection region over $s\in\{1\}\times[0.8,0.92]$ further shows that the null that $t$ is dilation-inducing is not rejected for $s_1\leq0.92$ at the $5\%$ significance level. The estimates reported across the 34 studies reviewed by \citet{arevalo2020false} suggest that \(s_1\) may plausibly be below \(0.9\). The median estimated false-negative rate is \(0.11\), corresponding to an implied true-positive rate of \(0.89\).

\section{Hypothesis Test Implementation}\label{app:computation}

This section describes the implementation of the test in \Cref{sect:inference}. We first simplify the tail probability in \eqref{eq:fixed-distribution-binomial-tail}. We then explain how branch-and-bound yields bounds on its supremum over $C_N\cap\mathcal P_{0,s}$.

If $H_n(N)=0$, every hypothetical table satisfies $H_n(n')\ge H_n(N)$, so $p_{\Vec p}(N)=1$ for every $\Vec p$. Suppose henceforth that $H_n(N)>0$. The factor $n^2(s_0+s_1-1)$ in \eqref{eq:exact-band-statistic} cancels the denominators in the sample gaps, allowing the statistic to be written directly in terms of counts. For each fixed pair $(n'_{\cdot1},n'_{10})$ with $n'_{\cdot1}\ne ns_1$, the statistic is the maximum of zero and two affine functions of $n'_{11}$ with opposite slopes. Consequently, the event $H_n(n')\ge H_n(N)$ is the union of a lower and an upper tail in $n'_{11}$.

Denote the integer cutoffs by $\ell(n'_{\cdot1},n'_{10})$ and $k(n'_{\cdot1},n'_{10})$. If $n'_{\cdot1}<ns_1$, expanding \eqref{eq:exact-band-statistic} gives
\begin{align*}
\ell(n'_{\cdot1},n'_{10})
&:=\left\lfloor
\frac{
(n'_{\cdot1}-n(1-s_0))(n'_{10}-n(1-s_1))-H_n(N)
}{
ns_1-n'_{\cdot1}
}
\right\rfloor,\\
k(n'_{\cdot1},n'_{10})
&:=\left\lceil
\frac{
(n'_{\cdot1}-n(1-s_0))n'_{10}
+n(1-s_0)(ns_1-n'_{\cdot1})+H_n(N)
}{
ns_1-n'_{\cdot1}
}
\right\rceil.
\end{align*}
If $n'_{\cdot1}>ns_1$, the two fractions exchange roles before taking the floor and ceiling. If $n'_{\cdot1}=ns_1$, the sample gaps equal
$n'_{00}/n$ and $n'_{10}/n$, respectively. Hence,
$H_n(n')=0$, and no hypothetical table with this reference-positive count satisfies
$H_n(n')\ge H_n(N)>0$. Such tables can therefore
be omitted from the tail calculation. When
$n'_{\cdot1}\ne ns_1$, the event $H_n(n')\ge H_n(N)$ is thus equivalent to
\[
n'_{11}\le\ell(n'_{\cdot1},n'_{10})
\quad\text{or}\quad
n'_{11}\ge k(n'_{\cdot1},n'_{10}).
\]
If the two tails overlap, their union is the entire binomial support, so the conditional probability is one.
Write $q:=p_{\cdot1}$, $u:=p_{11}/p_{\cdot1}$, and $v:=p_{10}/p_{\cdot0}$. By \eqref{eq:factorization},
$N'_{11}\mid N'_{\cdot1}=n'_{\cdot1}
\sim\operatorname{Bin}(n'_{\cdot1},u)$.
Since $N'_{11}$ and $N'_{10}$ are conditionally independent
given $N'_{\cdot1}$, this distribution is unchanged by
additionally conditioning on $N'_{10}=n'_{10}$. Let $F(\cdot;n'_{\cdot1},u)$ denote its cumulative distribution function. Therefore, \eqref{eq:fixed-distribution-binomial-tail} can equivalently be written as
\begin{align*}
p_{\Vec p}(N)
&=\sum_{\substack{n'_{\cdot1}=0\\n'_{\cdot1}\ne ns_1}}^n
\binom{n}{n'_{\cdot1}}
q^{n'_{\cdot1}}(1-q)^{n-n'_{\cdot1}}
\nonumber\\
&\quad\times\sum_{n'_{10}=0}^{n-n'_{\cdot1}}
\binom{n-n'_{\cdot1}}{n'_{10}}
v^{n'_{10}}(1-v)^{n-n'_{\cdot1}-n'_{10}}
\nonumber\\
&\quad\times\min\Bigl\{1,\,
F\bigl(\ell(n'_{\cdot1},n'_{10});n'_{\cdot1},u\bigr) +1-
F\bigl(k(n'_{\cdot1},n'_{10})-1;n'_{\cdot1},u\bigr)
\Bigr\}.
\end{align*}
The minimum accounts for possible overlap of the two tails. The cutoffs depend only on $N$, $s$, and the hypothetical pair $(n'_{\cdot1},n'_{10})$ and can be computed once. Replacing the innermost sum by binomial tail probabilities reduces the calculation from $\binom{n+3}{3}$ individual tables to at most $(n+1)(n+2)/2$ pairs.

We use a branch-and-bound algorithm to obtain lower and upper bounds on
\[
S_N:=\sup_{\Vec p\in C_N\cap\mathcal P_{0,s}}
p_{\Vec p}(N).
\]

The algorithm starts with the rectangle defined by the three
confidence intervals in \eqref{eq:confidence-localization},
restricted to the maintained range of $q$. Each point $(q,u,v)$
determines the cell probabilities
\[
\Vec p=\bigl((1-q)(1-v),\,q(1-u),\,(1-q)v,\,qu\bigr).
\]
We first describe the calculations for a given box and then
explain how the algorithm selects and subdivides boxes.
For each box
\[
B=[\underline q,\overline q]\times
[\underline u,\overline u]\times
[\underline v,\overline v]
\subseteq[1-s_0,s_1]\times[0,1]^2,
\]
interval arithmetic gives the bounds
\begin{align*}
\underline g_L(B)
&:=
\frac{
\underline q(s_1-\overline q)\underline u
-\displaystyle\max_{q\in\{\underline q,\overline q\}}
\Bigl\{(q+s_0-1)
\bigl[(1-\underline q)\overline v+s_1-1\bigr]\Bigr\}
}{s_0+s_1-1},
\\[6pt]
\overline g_L(B)
&:=
\frac{
\overline q(s_1-\underline q)\overline u
-\displaystyle\min_{q\in\{\underline q,\overline q\}}
\Bigl\{(q+s_0-1)
\bigl[(1-\overline q)\underline v+s_1-1\bigr]\Bigr\}
}{s_0+s_1-1},
\\[6pt]
\underline g_U(B)
&:=
\frac{
(1-s_0)(s_1-\overline q)
+(\underline q+s_0-1)(1-\overline q)\underline v
-\overline q(s_1-\underline q)\overline u
}{s_0+s_1-1},
\\[6pt]
\overline g_U(B)
&:=
\frac{
(1-s_0)(s_1-\underline q)
+(\overline q+s_0-1)(1-\underline q)\overline v
-\underline q(s_1-\overline q)\underline u
}{s_0+s_1-1}.
\end{align*}
These bounds satisfy
\[
\underline g_j(B)\le g_j(\Vec p;s)\le\overline g_j(B)
\qquad
\text{for all }(q,u,v)\in B,\quad j\in\{L,U\}.
\]
A box can therefore be discarded if
$\overline g_L(B)<0$ or $\overline g_U(B)<0$,
since it cannot contain a distribution satisfying both null inequalities.

For a retained box, evaluating the tail probability at a point in
$C_N\cap\mathcal P_{0,s}$ gives a lower bound on $S_N$. We search for
such a point by trying the midpoint and then the endpoints of the
$q$-interval and, for each choice of $q$, the midpoint and endpoints
of the $v$-interval. For fixed $q$ and $v$, both gap inequalities
are affine in $u$ and determine an interval of admissible values.
We intersect this interval with $[\underline u,\overline u]$.
If the intersection contains values satisfying the maintained
assumptions, we choose its midpoint and evaluate the tail probability
at the resulting distribution. The search within that box stops
at the first such point.

We initialize $\underline S_N=0$. For each box in which the search
finds a feasible point, we evaluate the tail probability at that
point and increase $\underline S_N$ if this value exceeds its
current value. Thus, $\underline S_N$ records the largest tail
probability evaluated at a feasible point across all boxes
processed so far. It is not reset when a box is subdivided.
Failure to find a feasible point among these trial values does not
establish that the box contains no null distribution, so it does
not justify discarding the box.

To obtain an upper bound on $S_N$, let
$A_{n'_{\cdot1}}(u,v)$ denote the tail probability conditional on
$N'_{\cdot1}=n'_{\cdot1}$. Then
\[
p_{\Vec p}(N)
=
\sum_{n'_{\cdot1}=0}^n
\binom{n}{n'_{\cdot1}}
q^{n'_{\cdot1}}(1-q)^{n-n'_{\cdot1}}
A_{n'_{\cdot1}}(u,v).
\]
For each fixed $n'_{\cdot1}\ne ns_1$, define the conditional
lower- and upper-tail probabilities by
\begin{align*}
A^-_{n'_{\cdot1}}(u,v)
&:=
\sum_{n'_{10}=0}^{n-n'_{\cdot1}}
\binom{n-n'_{\cdot1}}{n'_{10}}
v^{n'_{10}}(1-v)^{n-n'_{\cdot1}-n'_{10}}
\times
F\bigl(\ell(n'_{\cdot1},n'_{10});n'_{\cdot1},u\bigr),
\\[6pt]
A^+_{n'_{\cdot1}}(u,v)
&:=
\sum_{n'_{10}=0}^{n-n'_{\cdot1}}
\binom{n-n'_{\cdot1}}{n'_{10}}
v^{n'_{10}}(1-v)^{n-n'_{\cdot1}-n'_{10}}
\times
\Bigl[1-F\bigl(k(n'_{\cdot1},n'_{10})-1;
n'_{\cdot1},u\bigr)\Bigr].
\end{align*}
When $n'_{\cdot1}=ns_1$, we set both components to zero,
as these tables have $H_n(N')=0<H_n(N)$. Since the event defining
$A_{n'_{\cdot1}}(u,v)$ is their union,
\[
A_{n'_{\cdot1}}(u,v)
\le
\min\bigl\{1,\,
A^-_{n'_{\cdot1}}(u,v)+A^+_{n'_{\cdot1}}(u,v)
\bigr\}.
\]
We bound the two components separately over the box $B$.
Increasing $u$ shifts the conditional distribution of $N'_{11}$
toward larger counts. Thus, the lower-tail component is
nonincreasing in $u$, while the upper-tail component is
nondecreasing in $u$. Their upper bounds therefore use
$\underline u$ and $\overline u$, respectively.

The choice of endpoints for $v$ depends on how the cutoffs vary
with $n'_{10}$. Increasing $v$ shifts the conditional distribution
of $N'_{10}$ toward larger counts. If both $\ell$ and $k$ are
nondecreasing in $n'_{10}$, larger counts make the lower-tail
event easier to satisfy and the upper-tail event harder to
satisfy. Hence, the lower-tail component is nondecreasing in $v$,
while the upper-tail component is nonincreasing in $v$. This gives
\[
\overline A_{n'_{\cdot1}}(B)
:=
\min\Bigl\{1,\,
A^-_{n'_{\cdot1}}(\underline u,\overline v)
+
A^+_{n'_{\cdot1}}(\overline u,\underline v)
\Bigr\}.
\]
If both cutoffs are nonincreasing in $n'_{10}$, the directions
reverse, giving
\[
\overline A_{n'_{\cdot1}}(B)
:=
\min\Bigl\{1,\,
A^-_{n'_{\cdot1}}(\underline u,\underline v)
+
A^+_{n'_{\cdot1}}(\overline u,\overline v)
\Bigr\}.
\]
When $n'_{\cdot1}=ns_1$, we set
$\overline A_{n'_{\cdot1}}(B)=0$.
Each endpoint evaluation bounds its corresponding tail component
for every point in $B$. Adding these inequalities therefore gives
an upper bound on $A_{n'_{\cdot1}}(u,v)$ throughout the box, even
though the two components are evaluated at different points.
Since $A_{n'_{\cdot1}}(u,v)$ is a probability, we take the minimum
of this sum and one.

Substituting these conditional bounds into the mixture representation
gives, for every $(q,u,v)\in B$,
\[
p_{\Vec p}(N)
\le
\sum_{n'_{\cdot1}=0}^n
\binom{n}{n'_{\cdot1}}
q^{n'_{\cdot1}}(1-q)^{n-n'_{\cdot1}}
\overline A_{n'_{\cdot1}}(B).
\]
The right-hand side is a polynomial in $q$, expressed in the
Bernstein basis. We must bound this polynomial over the box's
interval $[\underline q,\overline q]$. For $\underline q<\overline q$, the same polynomial can be written as
\[
\sum_{n'_{\cdot1}=0}^n
\binom{n}{n'_{\cdot1}}
\left(\frac{q-\underline q}{\overline q-\underline q}\right)^{n'_{\cdot1}}
\left(\frac{\overline q-q}{\overline q-\underline q}\right)^{n-n'_{\cdot1}}
\widetilde A_{n'_{\cdot1}}(B),
\]
where the coefficients $\widetilde A_{n'_{\cdot1}}(B)$ are computed
from $\overline A_{n'_{\cdot1}}(B)$ by de Casteljau subdivision,
which repeatedly takes convex combinations of adjacent coefficients.
For a fixed box, these coefficients are constants; only the
weights multiplying them vary with $q$.
For every $q\in[\underline q,\overline q]$, the weights multiplying
these coefficients are nonnegative and sum to one. The polynomial
is therefore bounded above by its largest coefficient. This gives
the box's upper bound
\[
\overline S_N(B)
:=
\max_{0\le n'_{\cdot1}\le n}
\widetilde A_{n'_{\cdot1}}(B).
\]
If $\underline q=\overline q$, we instead evaluate the polynomial
at that single value.
Since the retained boxes cover $C_N\cap\mathcal P_{0,s}$, taking
the largest of their upper bounds gives
\[
\overline S_N
:=
\max_{\text{retained }B}\overline S_N(B),
\qquad
S_N\le\overline S_N.
\]
If no boxes remain, we set $\overline S_N=0$.

At each iteration, the global bounds satisfy
\[
\underline S_N\le S_N\le\overline S_N.
\]
The algorithm rejects if $\beta_n+\overline S_N\le\alpha$
and does not reject if $\beta_n+\underline S_N>\alpha$.
If neither condition holds and computation continues, it selects
the retained box with the largest upper bound $\overline S_N(B)$.
Within that box, it selects the longest of the three intervals
$[\underline q,\overline q]$,
$[\underline u,\overline u]$, and
$[\underline v,\overline v]$
and splits it at its midpoint. Keeping the other two intervals
unchanged gives two child boxes.

Both children are processed using the calculations above:
we check the gap bounds, search for a feasible point, and compute
an upper bound for each child that is not discarded.
The selected parent is then replaced by its retained children;
all other retained boxes remain in the search.

We recompute $\overline S_N$ as the largest upper bound among
all currently retained boxes. The lower bound $\underline S_N$
records the largest tail probability evaluated at any feasible
point found so far, including points found in earlier iterations.
We then repeat the comparison with $\alpha$.

Computation may also stop when the bounds are sufficiently close
or a prescribed work limit is reached. If
$\underline S_N\le\alpha-\beta_n<\overline S_N$,
the comparison remains unresolved. Rejection is based only on
the upper bound, preserving the finite-sample guarantee of
\Cref{th:finite-sample-validity}.

\section{Proofs
}\label{app:proofs}

\begin{proof}[Proof of \Cref{prop:post_test_updating_dilation}]
Fix any $c\geq0$ and any realized result $j\in\{0,1\}$. Since the test is
dilation-inducing, $\pi\in V_j(\Theta;\pi)$. Therefore, for every $d\in D$,
\[
    \inf_{p\in V_j(\Theta;\pi)}U(d,p,1)
    \leq
    U(d,\pi,1)
    \leq
    W^\pi.
\]
Taking the maximum over $d\in D$ gives
\[
    W_j
    =
    \max_{d\in D}\inf_{p\in V_j(\Theta;\pi)}U(d,p,1)
    \leq
    W^\pi.
\]
This proves the weak inequality for each result $j\in\{0,1\}$.

We next show that the inequality is strict for at least one result. Since the
test is dilation-inducing, $\{\pi\}\subsetneq V_1(\Theta;\pi)$. Hence, there
exists $\theta'\in\Theta$ such that
$v_1(\theta';\pi)\neq\pi$. If $v_1(\theta';\pi)>\pi$, set
$j'=1$ and $p_{j'}^+:=v_1(\theta';\pi)$. If instead
$v_1(\theta';\pi)<\pi$, then, by Bayes' theorem,
$\theta'_0+\theta'_1<1$, and therefore
$v_0(\theta';\pi)>\pi$. In this case, set
$j'=0$ and $p_{j'}^+:=v_0(\theta';\pi)$. Thus, in either case, there exists
$j'\in\{0,1\}$ and $p_{j'}^+\in V_{j'}(\Theta;\pi)$ such that
$p_{j'}^+>\pi$.

By the maintained health ordering, $h_0^d>h_1^d$ for each $d\in D$. Since $u$
is strictly increasing, $U(d,\cdot,1)$ is strictly decreasing for each
$d\in D$. Therefore, for every $d\in D$,
\[
    \inf_{p\in V_{j'}(\Theta;\pi)}U(d,p,1)
    \leq
    U(d,p_{j'}^+,1)
    <
    U(d,\pi,1)
    \leq
    W^\pi.
\]
Since $D$ is finite, taking the maximum over $d\in D$ preserves strict
inequality. Thus,
\[
    W_{j'}
    =
    \max_{d\in D}\inf_{p\in V_{j'}(\Theta;\pi)}U(d,p,1)
    <
    W^\pi.
\]
This proves the strict inequality for some result $j'\in\{0,1\}$.

Finally, suppose that $\pi\in\inter\big(V_k(\Theta;\pi)\big)$ for some
$k\in\{0,1\}$. Then there exist $\theta^-,\theta^+\in\Theta$ such that
\(
    v_k(\theta^-;\pi)<\pi<v_k(\theta^+;\pi).
\)
By Bayes' theorem, for any $\theta\in\Theta$,
\[
    v_1(\theta;\pi)>\pi
    \iff
    \theta_0+\theta_1>1
    \iff
    v_0(\theta;\pi)<\pi,
\]
and similarly
\[
    v_1(\theta;\pi)<\pi
    \iff
    \theta_0+\theta_1<1
    \iff
    v_0(\theta;\pi)>\pi.
\]
Thus, if one post-test set contains values strictly below and strictly above
$\pi$, then both post-test sets contain some value strictly above $\pi$. Hence,
for each realized result $j\in\{0,1\}$, there exists
$p_j^+\in V_j(\Theta;\pi)$ such that $p_j^+>\pi$.

Applying the same monotonicity argument as above, for every $d\in D$ and every
$j\in\{0,1\}$,
\[
    \inf_{p\in V_j(\Theta;\pi)}U(d,p,1)
    \leq
    U(d,p_j^+,1)
    <
    U(d,\pi,1)
    \leq
    W^\pi.
\]
Since $D$ is finite, taking the maximum over $d\in D$ preserves strict
inequality. So,
\[
    W_j
    =
    \max_{d\in D}\inf_{p\in V_j(\Theta;\pi)}U(d,p,1)
    <
    W^\pi.
\]
This proves the strict inequality for every result $j\in\{0,1\}$.
\end{proof}

\begin{proof}[Proof of \Cref{prop:dilation_characterization}]

We begin with a preliminary observation. By the definition of post-test probabilities and Bayes' theorem, for any pre-test probability $\pi\in(0,1)$ and each $j\in\{0,1\}$, $v_j(\theta;\pi)=\pi$ if and only if $\theta_0+\theta_1=1$. 

For sufficiency, assume that there are $\theta,\theta'\in\Theta$ such that $\theta_0+\theta_1=1$ and $\theta_0'+\theta_1'\neq1$, and fix an arbitrary pre-test probability $\pi\in(0,1)$. By the preliminary observation, $v_1(\theta;\pi)=v_0(\theta;\pi)=\pi$, so $\pi\in V_1(\Theta;\pi)$ and $\pi\in V_0(\Theta;\pi)$. Likewise, $v_1(\theta';\pi)\neq\pi$ and $v_0(\theta';\pi)\neq\pi$, so $V_1(\Theta;\pi)\neq\{\pi\}$ and $V_0(\Theta;\pi)\neq\{\pi\}$. Thus, $\{\pi\}\subsetneq V_1(\Theta;\pi)$ and $\{\pi\}\subsetneq V_0(\Theta;\pi)$.

For necessity, fix an arbitrary $\pi\in(0,1)$ and suppose that $\{\pi\}\subsetneq V_1(\Theta;\pi)$ and $\{\pi\}\subsetneq V_0(\Theta;\pi)$. Since $\pi\in V_1(\Theta;\pi)$, there exists $\theta\in\Theta$ such that $v_1(\theta;\pi)=\pi$, so $\theta_0+\theta_1=1$ by the preliminary observation. Since $V_1(\Theta;\pi)\neq\{\pi\}$, there exists $\theta'\in\Theta$ such that $v_1(\theta';\pi)\neq\pi$, so $\theta_0'+\theta_1'\neq1$.
\end{proof}

\begin{proof}[Proof of \Cref{th:simpler_inference}]
Under Assumptions \ref{assumption:reference_performance}
and \ref{assumption:prevalence},
\[
P_s(y=1)
=\frac{P(r=1)+s_0-1}{s_1+s_0-1}\in(0,1).
\]
For each $k\in\{0,1\}$, the marginal identity
\[
P(t=0,r=k)+P(t=1,r=k)
=P_s(r=k,y=0)+P_s(r=k,y=1)
\]
implies
\begin{align*}
&\min\{P(t=1,r=k),P_s(r=k,y=1)\}\\
&\quad-\max\{0,P(t=1,r=k)-P_s(r=k,y=0)\}\\
&=\min\{P(t=1,r=k),P(t=0,r=k),
         P_s(r=k,y=1),P_s(r=k,y=0)\}.
\end{align*}
Applying this equality to \eqref{eq:theta1L}
and \eqref{eq:theta1U} gives
\begin{align*}
&P_s(y=1)\big[\theta_1^U(s)-\theta_1^L(s)\big]\\
&=\min\{P(t=1,r=0),P(t=0,r=0),
        (1-s_1)P_s(y=1),s_0P_s(y=0)\}\\
&\quad+\min\{P(t=1,r=1),P(t=0,r=1),
        s_1P_s(y=1),(1-s_0)P_s(y=0)\}.
\end{align*}
Both terms are nonnegative, so
$\theta_1^L(s)\leq\theta_1^U(s)$.
Moreover, the definitions of the bounds imply
\begin{align*}
P_s(y=1)\theta_1^L(s)
&\geq \max\{0,P(t=1)-P_s(y=0)\},\\
P_s(y=1)\theta_1^U(s)
&\leq \min\{P(t=1),P_s(y=1)\}.
\end{align*}
These inequalities ensure that every
$\theta_1\in[\theta_1^L(s),\theta_1^U(s)]$
yields $(\theta_0,\theta_1)\in[0,1]^2$ through
the affine restriction in \eqref{eq:H_s}.
Thus, $\Theta_P(s)$ is nonempty and is a singleton
if and only if $\theta_1^L(s)=\theta_1^U(s)$.

If $s=(1,1)$, both terms in the width expression
are zero, so $\Theta_P(s)$ is a singleton.
If $s\neq(1,1)$, either $s_1<1$ or $s_0<1$.
Since $s_0+s_1>1$ implies $s_0,s_1>0$,
Assumptions \ref{assumption:prevalence}
and \ref{assumption:everything_goes} imply that
the first term is strictly positive when $s_1<1$,
and the second is strictly positive when $s_0<1$.
Consequently, $\Theta_P(s)$ is non-singleton
if and only if $s\neq(1,1)$.

Finally, the affine restriction in \eqref{eq:H_s}
implies that, for every $\theta\in\Theta_P(s)$,
\[
\theta_0+\theta_1
=1+\frac{\theta_1-P(t=1)}{P_s(y=0)}.
\]
This expression is strictly increasing in $\theta_1$,
so at most one point in $\Theta_P(s)$ satisfies
$\theta_0+\theta_1=1$.
Hence, $\Theta_P(s)$ contains both a point satisfying
$\theta_0+\theta_1=1$ and a point satisfying
$\theta_0+\theta_1\neq1$ if and only if it is
non-singleton and contains a point satisfying
$\theta_0+\theta_1=1$.
The result follows from
\Cref{prop:dilation_characterization}.
\end{proof}
    
\begin{proof}[Proof of \Cref{prop:primitive-dilation-null}]
The affine restriction defining $\Theta_P(s)$ is
\[
\theta_0
=
1+
\frac{\theta_1P_s(y=1)-p_{1\cdot}}{P_s(y=0)}.
\]
If $(\theta_0,\theta_1)$ is also on the antidiagonal, then
$\theta_0=1-\theta_1$. Hence,
\[
1-\theta_1
=
1+
\frac{\theta_1P_s(y=1)-p_{1\cdot}}{P_s(y=0)}.
\]
Because $P_s(y=0)+P_s(y=1)=1$, this equality holds if and only if
\(
\theta_1=p_{1\cdot}.
\)
The only possible intersection is therefore
\(
\theta^\star(P)=(1-p_{1\cdot},p_{1\cdot}).
\)
Because $p_{1\cdot}\in[0,1]$, this point belongs to $[0,1]^2$.

It remains to determine when
\(
\theta_1^L(s)
\le p_{1\cdot}
\le \theta_1^U(s).
\)
We use the identity
\begin{equation}
p_{\cdot1}
=
s_1P_s(y=1)
+(1-s_0)P_s(y=0).
\label{eq:reference-margin-identity}
\end{equation}

First consider the upper bound. Since $P_s(y=1)>0$,
\begin{align*}
p_{1\cdot}\le\theta_1^U(s)
\iff
P_s(y=1)p_{1\cdot}
\le{}&
\min\bigl\{p_{10},(1-s_1)P_s(y=1)\bigr\}
\\
&+
\min\bigl\{p_{11},s_1P_s(y=1)\bigr\}.
\end{align*}
The right-hand side equals
\[
\min\left\{
p_{1\cdot},
p_{10}+s_1P_s(y=1),
p_{11}+(1-s_1)P_s(y=1),
P_s(y=1)
\right\}.
\]
The inequalities involving the first and fourth entries are automatic because
\[
P_s(y=1)p_{1\cdot}\le p_{1\cdot}
\quad\text{and}\quad
P_s(y=1)p_{1\cdot}\le P_s(y=1).
\]
The inequality involving the third entry is equivalent to
\begin{align*}
P_s(y=1)p_{1\cdot}
&\le p_{11}+(1-s_1)P_s(y=1)
\\
&\iff
p_{11}-P_s(y=1)(p_{1\cdot}+s_1-1)\ge0
\\
&\iff
g_L(\Vec p;s)\ge0.
\end{align*}

The inequality involving the second entry is also automatic. It is equivalent
to
\[
p_{11}
\le
s_1P_s(y=1)+p_{1\cdot}P_s(y=0).
\]
If $p_{1\cdot}\le s_1$, then
\(
p_{11}
\le p_{1\cdot}
\le s_1P_s(y=1)+p_{1\cdot}P_s(y=0).
\)
If $p_{1\cdot}>s_1$, then $s_1>1-s_0$ implies
$p_{1\cdot}>1-s_0$. Therefore,
\begin{align*}
p_{11}
&\le p_{\cdot1} 
=
s_1P_s(y=1)+(1-s_0)P_s(y=0)
<
s_1P_s(y=1)+p_{1\cdot}P_s(y=0).
\end{align*}
It follows that
\begin{equation}
p_{1\cdot}\le\theta_1^U(s)
\iff
g_L(\Vec p;s)\ge0.
\label{eq:upper-bound-gap-equivalence}
\end{equation}

Now consider the lower bound. Since $P_s(y=1)>0$,
\begin{align*}
\theta_1^L(s)\le p_{1\cdot}
\iff
&\max\bigl\{0,p_{10}-s_0P_s(y=0)\bigr\}
\\
&+
\max\bigl\{0,p_{11}-(1-s_0)P_s(y=0)\bigr\}
\\
&\le P_s(y=1)p_{1\cdot}.
\end{align*}
The left-hand side equals
\[
\max\left\{
0,
p_{10}-s_0P_s(y=0),
p_{11}-(1-s_0)P_s(y=0),
p_{1\cdot}-P_s(y=0)
\right\}.
\]
The inequalities involving the first and fourth entries are automatic because
\(
0\le P_s(y=1)p_{1\cdot}
\)
and
\(
P_s(y=1)p_{1\cdot}
-\bigl(p_{1\cdot}-P_s(y=0)\bigr)
=P_s(y=0)(1-p_{1\cdot})
\geq 0.
\)
The inequality involving the third entry is equivalent to
\begin{align*}
p_{11}-(1-s_0)P_s(y=0)
&\le P_s(y=1)p_{1\cdot}
\\
&\iff
p_{01}\ge P_s(y=1)(s_1-p_{1\cdot})
\\
&\iff
g_U(\Vec p;s)\ge0,
\end{align*}
where the second equivalence uses \eqref{eq:reference-margin-identity}.

The inequality involving the second entry is also automatic. It is equivalent
to
\(
p_{11}\ge P_s(y=0)(p_{1\cdot}-s_0).
\)
If $p_{1\cdot}\le s_0$, the right-hand side is nonpositive, so the result
follows from $p_{11}\ge0$. Suppose instead that $p_{1\cdot}>s_0$.
Since $p_{00}\ge0$,
\(
p_{11}\ge p_{1\cdot}+p_{\cdot1}-1.
\)
Moreover,
\(
p_{1\cdot}+p_{\cdot1}-1
-P_s(y=0)(p_{1\cdot}-s_0)
=
P_s(y=1)(p_{1\cdot}+s_1-1)
>0\), where the strict inequality follows from $P_s(y=1)>0$, $p_{1\cdot}>s_0$, and
$s_0+s_1>1$. Therefore,
\begin{equation}
\theta_1^L(s)\le p_{1\cdot}
\iff
g_U(\Vec p;s)\ge0.
\label{eq:lower-bound-gap-equivalence}
\end{equation}

Combining \eqref{eq:upper-bound-gap-equivalence} and
\eqref{eq:lower-bound-gap-equivalence} gives
\[
\theta^\star(P)\in\Theta_P(s)
\iff
g_L(\Vec p;s)\ge0
\quad\text{and}\quad
g_U(\Vec p;s)\ge0.
\]
Because $s\ne(1,1)$, the final claim follows from
\Cref{th:simpler_inference}.
\end{proof}

\begin{proof}[Proof of \Cref{th:finite-sample-validity}]
Fix $\Vec p_0\in\mathcal P_{0,s}$. We first show that
$p_{\Vec p_0}(N)$ is superuniform. For $\gamma\in[0,1]$, suppose that there
exists a value in the support of $H_n(N)$ whose inclusive upper-tail
probability under $\Vec p_0$ is no greater than $\gamma$, and let
\[
h_\gamma
:=
\min\left\{
h\in\operatorname{supp}_{\Vec p_0}\bigl(H_n(N)\bigr):
P_{\Vec p_0}\{H_n(N)\ge h\}\le\gamma
\right\}.
\]
By monotonicity of the upper-tail probability,
\(
\{p_{\Vec p_0}(N)\le\gamma\}
=
\{H_n(N)\ge h_\gamma\}.
\)

Therefore,
\(
P_{\Vec p_0}\{p_{\Vec p_0}(N)\le\gamma\}
=
P_{\Vec p_0}\{H_n(N)\ge h_\gamma\}
\le\gamma.
\)
If no such support value exists, the event
$\{p_{\Vec p_0}(N)\le\gamma\}$ is empty, and the same inequality holds.

On the event $\{\Vec p_0\in C_N\}$, the set
$C_N\cap\mathcal P_{0,s}$ contains $\Vec p_0$, and hence
\[
\sup_{\Vec p\in C_N\cap\mathcal P_{0,s}}
p_{\Vec p}(N)
\ge p_{\Vec p_0}(N).
\]
The empty-set convention in \eqref{eq:berger-boos-pvalue} also implies
$p_{\mathrm{BB}}(N)\ge\beta_n$ for every realization. Thus, for
$\gamma<\beta_n$,
\[
P_{\Vec p_0}\{p_{\mathrm{BB}}(N)\le\gamma\}=0.
\]
For $\beta_n\le\gamma<1$, rejection at threshold $\gamma$ on
$\{\Vec p_0\in C_N\}$ implies
\(
p_{\Vec p_0}(N)\le\gamma-\beta_n.
\)
It follows that
\begin{align*}
P_{\Vec p_0}\{p_{\mathrm{BB}}(N)\le\gamma\}
&\le
P_{\Vec p_0}\{\Vec p_0\notin C_N\}
+
P_{\Vec p_0}
\{p_{\Vec p_0}(N)\le\gamma-\beta_n\}
\\
&\le
\beta_n+(\gamma-\beta_n)
=\gamma,
\end{align*}
where the second inequality follows from
\eqref{eq:conf_int} and the superuniformity of
$p_{\Vec p_0}(N)$. The result is immediate when $\gamma=1$. Since
$\Vec p_0\in\mathcal P_{0,s}$ was arbitrary, taking the supremum over the
null completes the proof.
\end{proof}

\begin{proof}[Proof of \Cref{prop:dilation_characterization_set}]
We begin with a preliminary observation. By the definition of post-test probabilities and Bayes' theorem, for any pre-test probability $\pi\in(0,1)$ and each $j\in\{0,1\}$, $v_j(\theta;\pi)=\pi$ if and only if $\theta_0+\theta_1=1$. Moreover, if $\theta_0+\theta_1>1$, then $v_1(\theta;\pi)>\pi>v_0(\theta;\pi)$, while if $\theta_0+\theta_1<1$, then $v_1(\theta;\pi)<\pi<v_0(\theta;\pi)$.

For sufficiency, assume that there are $\theta,\theta'\in\Theta$ such that $\theta_0+\theta_1=1$ and $\theta_0'+\theta_1'\neq1$, and fix an arbitrary closed nonempty $\Pi\subseteq(0,1)$. By the preliminary observation, $v_1(\theta;\pi)=v_0(\theta;\pi)=\pi$ for every $\pi\in\Pi$, so $\Pi\subseteq V_1(\Theta;\Pi)$ and $\Pi\subseteq V_0(\Theta;\Pi)$. Since $\Pi$ is closed and bounded, let $\underline\pi:=\min\Pi$ and $\overline\pi:=\max\Pi$. If $\theta_0'+\theta_1'>1$, then $v_1(\theta';\overline\pi)>\overline\pi$ and $v_0(\theta';\underline\pi)<\underline\pi$, so $V_1(\Theta;\Pi)\neq\Pi$ and $V_0(\Theta;\Pi)\neq\Pi$. If $\theta_0'+\theta_1'<1$, then $v_1(\theta';\underline\pi)<\underline\pi$ and $v_0(\theta';\overline\pi)>\overline\pi$, so likewise $V_1(\Theta;\Pi)\neq\Pi$ and $V_0(\Theta;\Pi)\neq\Pi$. Thus, $\Pi\subsetneq V_1(\Theta;\Pi)$ and $\Pi\subsetneq V_0(\Theta;\Pi)$.

For necessity, fix an arbitrary $\pi\in(0,1)$ and let $\Pi=\{\pi\}$. Since the index test is dilation-inducing, $\{\pi\}\subsetneq V_1(\Theta;\{\pi\})$ and $\{\pi\}\subsetneq V_0(\Theta;\{\pi\})$. Since $\pi\in V_1(\Theta;\{\pi\})$, there exists $\theta\in\Theta$ such that $v_1(\theta;\pi)=\pi$, so $\theta_0+\theta_1=1$ by the preliminary observation. Since $V_1(\Theta;\{\pi\})\neq\{\pi\}$, there exists $\theta'\in\Theta$ such that $v_1(\theta';\pi)\neq\pi$, so $\theta_0'+\theta_1'\neq1$.
\end{proof}

\begin{proof}[Proof of \Cref{prop:primitive-dilation-nullS}]
For each $s\in\Tilde{\mathcal{S}}$, the affine restriction in \eqref{eq:H_s} can be written as
\[
p_{1\cdot}
=
\theta_1 P_s(y=1)
+
(1-\theta_0)P_s(y=0).
\]
If $\theta_0+\theta_1=1$, this yields
\(
p_{1\cdot}
=
\theta_1\bigl[P_s(y=1)+P_s(y=0)\bigr]
=
\theta_1.
\)
Thus, every $\Theta_P(s)$, and hence their union $\Theta_P(\mathcal{S})$, can intersect the antidiagonal only at $\theta^\star(P)=(1-p_{1\cdot},p_{1\cdot})$.

For each $s\in\Tilde{\mathcal{S}}\setminus\{(1,1)\}$, \Cref{prop:primitive-dilation-null} gives
\[
\theta^\star(P)\in\Theta_P(s)
\iff
g_L(\Vec p;s)\ge0
\quad\text{and}\quad
g_U(\Vec p;s)\ge0.
\]
This equivalence also holds when $s=(1,1)$. Indeed, in that case,
\[
\Theta_P(s)
=
\left\{
\left(
\frac{p_{00}}{p_{\cdot0}},
\frac{p_{11}}{p_{\cdot1}}
\right)
\right\},
\qquad
g_L(\Vec p;s)
=
p_{11}-p_{\cdot1}p_{1\cdot},
\qquad
g_U(\Vec p;s)
=
-g_L(\Vec p;s).
\]
Both gap inequalities therefore hold if and only if
$p_{11}=p_{\cdot1}p_{1\cdot}$, which is precisely the condition that this singleton equals $\{\theta^\star(P)\}$. Taking the union over $s\in\Tilde{\mathcal{S}}$ proves the stated membership equivalence.

Finally, by the argument in the proof of \Cref{th:simpler_inference}, full support implies that $\Theta_P(s)$ is non-singleton for every feasible $s\neq(1,1)$. Since $\Tilde{\mathcal{S}}\setminus\{(1,1)\}\neq\emptyset$, the union $\Theta_P(\mathcal{S})$ is therefore non-singleton. Its only possible point on the antidiagonal is $\theta^\star(P)$, so \Cref{prop:dilation_characterization} implies that $t$ is dilation-inducing if and only if $\theta^\star(P)\in\Theta_P(\mathcal{S})$. The conclusion follows.
\end{proof}

\end{document}